\documentclass[10pt,twocolumn,secnumarabic,amssymb, nobibnotes, aps, prd,superscriptaddress]{revtex4-2}
\usepackage{xcolor,soul}
\sethlcolor{red}
\usepackage{amsmath}
\usepackage{wrapfig}
\usepackage[utf8]{inputenc}

\usepackage{lipsum}
\usepackage{subfigure}
\usepackage{wrapfig}
\usepackage{xparse,xcoffins}
\usepackage{float}

\ExplSyntaxOn
\NewCoffin\imagecoffin
\NewCoffin\labelcoffin

\keys_define:nn { miguel/label }
 {
  label   .tl_set:N = \l_miguel_label_tl,
  labelbox .bool_set:N = \l_miguel_label_box_bool,
  labelbox .default:n = true,
  fontsize .tl_set:N = \l_miguel_label_size_tl,
  fontsize .initial:n = \footnotesize,
  pos .choice:,
  pos/nw .code:n = \tl_set:Nn \l_miguel_label_pos_tl { left,up },
  pos/ne .code:n = \tl_set:Nn \l_miguel_label_pos_tl { right,up },
  pos/sw .code:n = \tl_set:Nn \l_miguel_label_pos_tl { left,down },
  pos/se .code:n = \tl_set:Nn \l_miguel_label_pos_tl { right,down },
  pos/n .code:n = \tl_set:Nn \l_miguel_label_pos_tl { hc,up },
  pos/w .code:n = \tl_set:Nn \l_miguel_label_pos_tl { left,vc },
  pos/s .code:n = \tl_set:Nn \l_miguel_label_pos_tl { hc,down },
  pos/e .code:n = \tl_set:Nn \l_miguel_label_pos_tl { right,vc },
  pos .initial:n = nw,
  unknown .code:n   = \clist_put_right:Nx \l_miguel_label_clist
                       { \l_keys_key_tl = \exp_not:n { #1 } }
 }
\clist_new:N \l_miguel_label_clist
\box_new:N \l_miguel_label_box
\box_new:N \l_miguel_label_image_box 

\NewDocumentCommand{\xincludegraphics}{O{}m}
 {
  \group_begin:
  \tl_clear:N \l_miguel_label_tl
  \clist_clear:N \l_miguel_label_clist
  \keys_set:nn { miguel/label } { #1 }
  \tl_if_empty:NTF \l_miguel_label_tl
   {
    \miguel_includegraphics:Vn \l_miguel_label_clist { #2 }
   }
   {
    \SetHorizontalCoffin\imagecoffin
     {
      \miguel_includegraphics:Vn \l_miguel_label_clist { #2 }
     }
    \SetHorizontalCoffin\labelcoffin
     {
      \raisebox{\depth}
       {
        \bool_if:NTF \l_miguel_label_box_bool
         { \fcolorbox{white}{white}{\l_miguel_label_size_tl\l_miguel_label_tl} }
         { \l_miguel_label_size_tl\l_miguel_label_tl }
       }
     }
    \SetVerticalPole\imagecoffin{left}{3pt+\CoffinWidth\labelcoffin/2}
    \SetVerticalPole\imagecoffin{right}{\Width-3pt-\CoffinWidth\labelcoffin/2}
    \SetHorizontalPole\imagecoffin{up}{\Height-3pt-\CoffinHeight\labelcoffin/2}
    \SetHorizontalPole\imagecoffin{down}{3pt+\CoffinHeight\labelcoffin/2}
    \use:x{\JoinCoffins\imagecoffin[\l_miguel_label_pos_tl]\labelcoffin[vc,hc]}
    \TypesetCoffin\imagecoffin
   }
   \group_end:
 }
\NewDocumentCommand{\setlabel}{m}
 {
  \keys_set:nn { miguel/label } { #1 }
 }

\cs_new_protected:Nn \miguel_includegraphics:nn
 {
  \includegraphics[#1]{#2}
 }
\cs_generate_variant:Nn \miguel_includegraphics:nn { V }

\ExplSyntaxOff

\makeatletter
\newcommand{\twocolumncaption}{\@dblarg\@twocolumncaption}
\def\@twocolumncaption[#1]#2{%
  \renewcommand{\@makecaption}[2]{%
    \par\vskip\abovecaptionskip\begingroup\small\rmfamily
    \splittopskip=0pt
    \setbox\@tempboxa=\vbox{
      \@arrayparboxrestore \let \\\@normalcr
      \hsize=.5\hsize \advance\hsize-1em
      \let\\\heading@cr
      \noindent ##1\ ##2\par
    }%
    \vbadness=10000
    \setbox\z@=\vsplit\@tempboxa to .55\ht\@tempboxa
    \setbox\z@=\vtop{\hrule height 0pt \unvbox\z@}
    \setbox\tw@=\vtop{\hrule height 0pt \unvbox\@tempboxa}
    \noindent\box\z@\hfill\box\tw@\par
    \endgroup\vskip \belowcaptionskip
  }%
  \setlength{\abovecaptionskip}{4ex}%
  \caption[#1]{#2}%
}

\usepackage{graphicx}
\begin{document}

\title{Influence of helicoidal spin-orbit coupling and Rabi coupling in  dynamics of 2D Bose-Einstein Condensates}%

\author{Sai Satyaprakash Biswal}%
\email {saisatyaprakash2002@gmail.com}
\author{S. Saravana Veni}%
\email {s\textunderscore saravanaveni@cb.amrita.edu.in}
\affiliation{Department of Physics, Amrita School of Physical Sciences, Amrita Vishwa Vidyapeetham, Coimbatore, 641112, Tamil Nadu, India}

\begin{abstract}
This study explores the dynamics of Bose-Einstein condensates (BECs) with helicoidal spin-orbit coupling (SOC) and Rabi coupling, confined under a two-dimensional harmonic potential. The relationship between helicoidal SOC, non-linear interactions along with Rabi coupling and their impact on the stability and non linear trapped modes of the condensate is analyzed using a coupled Gross-Pitaevski (GP) framework. A linearized GP equation is used to investigate modulation instability (MI), demonstrating the impact of strong coupling effects and anisotropic confinement on the instability dynamics. It has been shown that the modulation instability in the condensate is predominantly governed by the competition between the attractive and repulsive mean-field interactions. Additionally, stability regimes are altered by the harmonic confinement, enhancing their susceptibility to SOC-induced asymmetry and intra- and intercomponent interactions. These results shed light on the possibility for unique quantum phases and emergent characteristics of helicoidal SOC-driven condensates.

\end{abstract}
\maketitle

\section{Introduction}

In Bose-Einstein Condensates (BECs), modulation instability (MI) is a nonlinear phenomenon in which minor perturbations cause a spatially or temporally uniform state to become unstable, thus causing the perturbations to rise exponentially \cite{ref1}. This instability often leads to the creation of localized structures like solitons, vortices, or patterns, depending upon the initial conditions and systems parameters \cite{ref2}. Goldstein and Meystre were the first to study modulation instability (MI) in two-component BECs \cite{ref4} and they demonstrated that MI can happen even in systems with repulsive interactions \cite{ref5,ref6}. This counterintuitive behaviour, which arises due to the inter-component coupling, causes disturbances to develop exponentially and spatial patterns to appear. Over the recent decades, the concept of MI has been extended to one- and two-component Bose-Einstein condensates with modified Gross-Pitaevskii equations (GPEs), which have been improved to rightly characterize and describe the presence of matter-waves \cite{ref7,ref8}. MI has been examined across various fields and each has shown its impact on complex systems. It has been observed that it affects the turbulence and wave production in nonlinear fluids \cite{ref9}, magnetic domain formation in condensed matter systems \cite{ref10}, and turbulence and wave motion in plasmas \cite{ref11}. In BECs, MI is critical for the understanding of superfluid dynamics and soliton formation \cite{ref12,ref13,ref14}. Thus, each field highlights MI’s role in nonlinear wave phenomena. To investigate the effect of spin-orbit interaction on the dynamics of the instability, spin-orbit coupling (SOC) has been applied to MI in BECs \cite{ref15}. SOC describes the interplay between a particle’s spin and orbital momentum and has been a key concept in semiconductor physics as well, bridging our understanding about phenomena like the spin-Hall effect \cite{ref16}, topological insulators \cite{ref17}, and the emergence of spintronics \cite{ref18,ref19}. Examining MI in BECs is essential because it leads to the creation of spatially localized patterns in the form of self-bound quantum droplets (QDs) and bright solitons. Through the recent theoretical studies of binary BECs it can be easily understood that MI results in the production of many QDs, each of which has its own distinct properties \cite{ref20,ref21}. Bose-Einstein condensates (BECs) can be described by their localized density distributions under nontrivial interactions \cite{ref22}. Rich dynamical features, including finite-density peaks that contrast with the variable peak densities seen in typical bright solitons, are present in condensates subject to helicoidal spin–orbit coupling. The number of atoms and the interaction between repulsive and attractive mean-field interactions have a significant impact on the condensate’s density response \cite{ref23,ref24,ref25}. The condensate density tends to saturate in regimes with enough atoms, indicating an effective incompressibility that resembles some collective fluid-like properties \cite{ref26,ref27}. These behaviors demonstrate how condensate dynamics can lead to interesting instabilities and pattern formation, which are of fundamental theoretical interest, particularly under helicoidal SOC \cite{ref28,ref29}. The production of multicomponent condensates, where the atoms are trapped in several internal states inside the same trapping potential, has been made easier with the help of the recent developments around trapping techniques \cite{ref30,ref31}. Not only do these systems offer a flexible framework for investigating complicated processes in BECs, but they also take various quantum states into consideration. In multicomponent species, other interactions such as interspecies interactions between different components can also occur and it is essential to take them into account. These interactions often lead to rich and intricate dynamical phenomena, including phase separations, coupled excitations, and pattern formations \cite{ref32,ref33,ref34}. These phenomena help us draw a better picture of the interactions of quantum states in ultracold systems as they are not present in single-component condensates. Recent progress in synthetic gauge fields and Raman-dressed spin–orbit coupling has significantly expanded the range of experimentally accessible SOC geometries in ultracold atomic gases.  A number of schemes have been put forth to extend standard Rashba- or Dresselhaus-type interactions toward more complex geometries since the groundbreaking discovery of SOC in BECs \cite{ref35,ref36}. 
{Although helicoidal SOC has not yet been directly realized in ultracold atomic experiments, several theoretical proposals indicate that such couplings can be engineered in optical systems, for example, in helical waveguide arrays \cite{rechtsman2013photonic}. The development of synthetic gauge fields in ultracold atomic systems was reported in Ref. \cite{lin2011spin}, demonstrating the coupling between neutral atoms and externally engineered gauge potentials. Building on these advances, SOC in BECs can be experimentally tuned and, in principle, configured in a helicoidal form \cite{ref3}. More recently, the modulation instability in the presence of helicoidal SOC has been investigated in Ref. \cite{otlaadisa2021modulation}. These schemes enable controlled generation of spatially twisted (helicoidal) spin–momentum coupling while preserving tunability of both the SOC strength and the Rabi coupling. Our analysis is therefore experimentally motivated and lies within parameter regimes accessible to current cold-atom platforms.}
Accessible methods for implementing helicoidal gauge fields, specifically optical Raman coupling and Floquet-engineered lattice potentials, preserve the tunability of SOC strength and Rabi frequency within experimentally achievable ranges.  Realizations in $^{87}$Rb condensates with $10^4$--$10^5$ atoms \cite{ref37} are consistent with typical SOC strengths on the order of a few recoil momenta and Rabi couplings in the kilohertz regime. The practical relevance of the current theoretical analysis is further highlighted by the observation of supersolid stripe phases in spin--orbit coupled condensates \cite{ref38}, which shows that current experiments can probe regimes where helicoidal SOC effects may manifest. Although MI in BECs has been studied under different SO couplings, this system has not been thoroughly examined when restricted within a real-world potential trap such as the harmonic potential \cite{ref39,ref40,ref41}. Such studies have frequently been limited to uni-dimensional analysis as well but in this paper, we have attempted to include the condensates under a harmonic potential in 2D \cite{ref42, ref43, ref44, ref45,ref46,ref47}. We further investigate the MI in BECs with two-component helicoidal SO-coupling which are equally distributed between the pseudospin states and contained within a 2D harmonic potential. A small perturbation approximation has been used to obtain a linearised Gross-Pitaevskii (GP) equation and the dispersion relation associated with the instability of flat continuous wave (cw) basis. We provide comprehensive assessments of how MI in two-component BECs is affected by the helicoidal gauge potential, SO coupling, nonlinear interactions and 2D harmonic confinement. We establish that anisotropic confinement effects introduced by the 2D harmonic potential alter the dynamics of MI and have a significant impact on the interaction between the helicoidal SOC and the nonlinear interactions as well. Despite violating the symmetry characteristics of MI and showing a strongly linked response with SO coupling and intra- and intercomponent atomic interactions, the helicoidal gauge potential is still a vital aspect. Additionally, newer stability regimes are also introduced by the 2D harmonic potential which enhances the system’s susceptibility to these interactions. The make-up of the paper is as follows. The theoretical model of the two-component BECs with helicoidal SOC under 2D harmonic potential has been described in Sec.~2. The dispersion relation of the MI has also been derived using linear-stability analysis in this section. Sec.~3 elaborates upon a systematic discussion and the observations of the effects of the atomic interactions, helicoidal gauge potential, SO coupling and 2D harmonic confinement on MI. Simulations of this system which validate our theoretical prediction have also been described in this section and are followed by trapping frequency and condensate configurations in Sec.~4. This paper is concluded in Sec.~5.

\section{The Model}

We consider the model Hamiltonian associated with the following energy functional, which encapsulates the dynamics of a coupled two-component condensate subjected to spin-orbit coupling, nonlinear interactions, harmonic confinement, and linear Rabi coupling.

\begin{equation}
\begin{aligned}
\mathcal{H} =\ & 
\sum_{j=1}^{2} \left[ 
\frac{1}{2} |\nabla \psi_j|^2 
+ \frac{g_j}{2} |\psi_j|^4 
+ \frac{\omega^2}{2}(x^2 + y^2) |\psi_j|^2 
\right] \\&\quad
+ g_{12} |\psi_1|^2 |\psi_2|^2 + R\, \mathrm{Re}(\psi_1^* \psi_2) \\
&+ i \alpha \left[ \psi_1^* \left( \partial_x + \partial_y \right) \psi_2 
- \psi_2^* \left( \partial_x + \partial_y \right) \psi_1 \right] \\
&- i \beta \left[ \psi_1^* \left( \partial_x + \partial_y \right) \psi_1 
+ \psi_2^* \left( \partial_x + \partial_y \right) \psi_2 \right]
\end{aligned}
\end{equation}

The wavefunctions \(\psi_1\) and \(\psi_2\) represent the two pseudo-spin components for the coupled two-component BEC, in the energy functional above. Each term in the functional has an explained purpose. For example, \(g_1\) and \(g_2\) represent the intra-species interaction strength resulting from contact \(s\)-wave scattering. The term \(\frac{1}{2}|\nabla \psi_j|^2\) represents the kinetic energy for both components. Inter-species interactions are described by \(g_{12}|\psi_1|^2|\psi_2|^2\) as the nonlinear coupling between the two components. \(\frac{1}{2} \omega^2 (x^2 + y^2)\) constitutes the harmonic trapping potential, where \(\omega\) is the trap frequency needed for symmetric confinement in the two-dimensional plane. The Rabi coupling strength term \(R\) determines the coherent linear interconversion between \(\psi_1\) and \(\psi_2\) and is most commonly induced by an external Raman process. The helicoidal SOC is introduced by the term with coefficient \(\alpha\), which couples spatial derivatives to spin degrees and allows momentum-dependent spin mixing. The parameter \(\beta\) describes helicoidal self-coupling (or diagonal SOC-like interactions), introducing momentum-dependent effects independently on each component. After appropriate rescaling according to the inherent length and energy scales of the system, all parameters appear in dimensionless form. The combination of all these terms sustains a rich dynamical landscape that strongly influences the condensate’s stability and modulation instability behavior.

A dimensional form of the coupled Gross--Pitaevskii equations with helicoid
(spin-orbit type) and Rabi couplings under a two-dimensional harmonic trap is
\begin{eqnarray}
i\hbar\,\partial_t \Psi_1 &=& -\frac{\hbar^2}{2m}\nabla^2\Psi_1
- i\hbar(\partial_x+\partial_y)\big(\alpha_{0}\,\Psi_2 - \beta_{0}\,\Psi_1\big) \nonumber\\
&&+ \Big(G_{11}|\Psi_1|^2 + G_{12}|\Psi_2|^2\Big)\Psi_1 \\
&& + \tfrac12 m(\omega_x^2 x^2+ \omega_y^2 y^2)\Psi_1\nonumber + \frac{\hbar R_{0}}{2}\,\Psi_2, \nonumber
\label{gp1}
\end{eqnarray}
\begin{eqnarray}
i\hbar\,\partial_t \Psi_2 &=& -\frac{\hbar^2}{2m}\nabla^2\Psi_2
- i\hbar(\partial_x+\partial_y)\big(\alpha_{0}\,\Psi_1 + \beta_{0}\,\Psi_2\big)\nonumber\\
&&+ \Big(G_{22}|\Psi_2|^2 + G_{12}|\Psi_1|^2\Big)\Psi_2
\\
&& + \tfrac12 m(\omega_x^2 x^2+\omega_y^2 y^2)\Psi_2\nonumber + \frac{\hbar R_{0}}{2}\,\Psi_1,\nonumber
\label{gp2}
\end{eqnarray} 
with $\nabla^2=\partial_x^2+\partial_y^2$. Here $\alpha,\beta$ parameterize the helicoid (linear-derivative) couplings, $G_{ij}$ the 2D interaction strengths, $\omega_{x}$, $\omega_{y}$ are the frequency along the x and y axis and  $R$ denotes the Rabi frequency.
The dimensionaless form of the GP equation is given by 
\begin{eqnarray}
i \frac{\partial \psi_1}{\partial t} &=& -\frac{1}{2} \nabla^2 \psi_1 - i \left( \frac{\partial}{\partial x} + \frac{\partial}{\partial y} \right) \left(  \alpha \psi_2 - \beta \psi_1 \right)\nonumber \\
&&+ \left( g_1 |\psi_1|^2 + g_{12} |\psi_2|^2 \right) \psi_1 
+ \frac{\omega^2}{2}(x^2 + y^2) \psi_1 \nonumber\\
&&+ \frac{R}{2} \psi_2
\label{dgp1}
\end{eqnarray}

\begin{eqnarray}
i \frac{\partial \psi_2}{\partial t} &=& -\frac{1}{2} \nabla^2 \psi_2 - i \left( \frac{\partial}{\partial x} + \frac{\partial}{\partial y} \right) \left(  \alpha \psi_1 + \beta \psi_2 \right)\nonumber \\
&&+ \left( g_2 |\psi_2|^2 + g_{12} |\psi_1|^2 \right) \psi_2 + \frac{\omega^2}{2}(x^2 + y^2) \psi_2 \nonumber \\
&&+ \frac{R}{2} \psi_1
\label{dgp2}
\end{eqnarray}

The dimensionless equations are obtained by choosing a reference frequency 
$\omega_0$ (e.g.~$\omega_0=\sqrt{\omega_x\omega_y}$) and defining length in unit of 
$\ell_0 = \sqrt{\frac{\hbar}{m\omega_0}}$, time in units of
$t_0 = \omega_0^{-1}$ and  energy in terms of 
$E_0 = \hbar \omega_0$ with the dimensionless parameters
\begin{eqnarray}
\alpha &=& \frac{\alpha_{0}}{\ell_0 \omega_0}, \qquad 
\beta  = \frac{\beta_{0}}{\ell_0 \omega_0}, \qquad
g_{jk} = \frac{G_{jk}}{\hbar \omega_0 \ell_0^2}, \qquad\nonumber\\
&R&= \frac{R_{0}}{\omega_0}, \qquad
\omega = \frac{\omega_{x,y}}{\omega_0}. \nonumber
\end{eqnarray}
such that the normalization is
\begin{eqnarray}
N = \int dx\,dy \left( |\psi_1|^2 + |\psi_2|^2 \right) = N_1 + N_2. 
\label{N}
\end{eqnarray}
Here we set $N_{1}=N_{2}=1$. $\alpha$ depicts the spin-orbit (SO) coupling strength and $\beta$ constitutes the helicoidal gauge potential.

The stationary solutions of Eq. (4) and Eq. (5) are obtained using the ansatz 
\[begin{eqnarray}]
\psi_j = e^{-i\mu t} \sqrt{n_{j0}}
\label{psi}
\]
,with homogeneous background densities $n_{10}$ and $n_{20}$ and chemical potential $\mu$. The expression for the chemical potential $\mu$ is given by,
\begin{eqnarray}
\mu &=& \int d^2 r \Bigg\{ 
\sum_{j=1}^2 \Big[ \tfrac{1}{2}|\nabla\psi_j|^2 
+ \tfrac{1}{2}\big(\lambda_x^2  x^2 + \lambda_y^2  y^2\big)|\psi_j|^2\nonumber\\
&&- i\,\psi_j^*\big(\alpha\,\partial_{ x} + \beta\,\partial_{ y}\big)\psi_j \Big] 
+ \tfrac{g_{11}}{2}|\phi_1|^4 
+ \tfrac{g_{22}}{2}|\phi_2|^4 \nonumber\\
&&+ g_{12}|\phi_1|^2|\phi_2|^2 
+ \tfrac{R}{2}\big(\phi_1^*\phi_2 + \phi_2^*\phi_1\big) 
\Bigg\}.
\label{mu}
\end{eqnarray}

To analyze and assess this system using MI, we have included the small perturbations $\delta \psi_j$ ($\delta \psi_j \ll \sqrt{n_{j0}}$) to the continuous wave (cw) solutions as follows: 
\begin{eqnarray}
\psi_j = e^{-i \mu t} \left( \sqrt{n_{j0}} + \delta \psi_j \right).
\end{eqnarray}
Below the linearized equations for the small perturbations have been shown,

\begin{equation}
\begin{aligned}
i \frac{\partial (\delta \psi_1)}{\partial t} &= 
- \frac{1}{2} \nabla^2 \delta \psi_1 - i \left( \frac{\partial}{\partial x} + \frac{\partial}{\partial y} \right) \left(  \alpha \delta \psi_1 - \beta \delta \psi_2 \right) \\
&\quad 
+ g_1 n_{10} (\delta \psi_1 + \delta \psi_1^*) 
+ g_{12} \sqrt{n_{10} n_{20}} (\delta \psi_2 + \delta \psi_2^*) \\
&\quad 
+ \frac{\omega^2}{2} (x^2 + y^2) \delta \psi_1 
+ \frac{R}{2} \delta \psi_2,
\end{aligned}
\end{equation}

\begin{eqnarray}
i \frac{\partial (\delta \psi_2)}{\partial t} &=& - \frac{1}{2} \nabla^2 \delta \psi_2 - i \left( \frac{\partial}{\partial x} + \frac{\partial}{\partial y} \right) \left(  \alpha \delta \psi_2 - \beta \delta \psi_1 \right) \nonumber\\
&&+ g_2 n_{20} (\delta \psi_2 + \delta \psi_2^*) + g_{12} \sqrt{n_{10} n_{20}} (\delta \psi_1 + \delta \psi_1^*) \nonumber\\
&&+ \frac{\omega^2}{2} (x^2 + y^2) \delta \psi_2 + \frac{R}{2} \delta \psi_1.
\end{eqnarray}

\smallskip

We further assume the solution for the perturbations in the form of plane waves as follows,

\begin{equation}
\begin{aligned}
\delta \psi_j &= \zeta_j \cos (K_x x + K_y y - \Omega t)
+ i \eta_j \sin (K_x x + K_y y - \Omega t), \\
&\qquad j = 1, 2 .
\end{aligned}
\end{equation}
\smallskip

Here, \(K_x\) and \(K_y\) represent the wave numbers along the \(x\)- and \(y\)-directions, respectively. \(\Omega\) is the complex eigen-frequency, and \(\zeta_j\) and \(\eta_j\) are the perturbation amplitudes. Substituting Eq.~(11) into Eqs.~(9) and (10) results a 4x4 matrix containing all the coefficients of \(\zeta_1\), \(\zeta_2\), \(\eta_1\) and \(\eta_2\).

\begin{equation}
M \times (\zeta_1, \zeta_2, \eta_1, \eta_2) = 0
\end{equation}
The matrix M along with its matrix elements can be found in the Appendix, later in this paper. 

To obtain nontrivial solutions upon solving for the det M = 0, we will obtain the dispersion relation for this system at eigen-frequency \(\Omega\) as,
\begin{equation}
\Omega^4 + P_3 \Omega^3 + P_2 \Omega^2 + P_1 \Omega + P_0 = 0
\end{equation}

Here in the above equation, the general expressions for the coefficients \(P_j\) (j =1, 2 and 3) have been given in the Appendix and can be solved systematically to get the solutions as mentioned below.

\begin{equation}
\begin{aligned}
\Omega_{1,2} &= -\frac{P_3}{4} - \frac{1}{2} \sqrt{ -\frac{2P_2}{3} + \frac{P_3^2}{4} + \Lambda } \\
&\quad \pm \frac{1}{2} \sqrt{ -\frac{4}{3}P_2 + \frac{P_3^2}{2} - \Lambda - 
\frac{8P_1 + 4P_2P_3 - P_3^2}{4 \sqrt{ -\frac{2P_2}{3} + \frac{P_3^2}{4} + \Lambda }} }
\end{aligned}
\end{equation}

\begin{equation}
\begin{aligned}
\Omega_{3,4} &= -\frac{P_3}{4} + \frac{1}{2} \sqrt{ -\frac{2P_2}{3} + \frac{P_3^2}{4} + \Lambda } \\
&\quad \pm \frac{1}{2} \sqrt{ -\frac{4}{3}P_2 + \frac{P_3^2}{2} - \Lambda + 
\frac{8P_1 + 4P_2P_3 - P_3^2}{4 \sqrt{ -\frac{2P_2}{3} + \frac{P_3^2}{4} + \Lambda }} }
\end{aligned}
\end{equation}
\smallskip

In the above equations, the expression for \(\Lambda\) can be written as,
\begin{equation}
\Lambda = \frac{2^{1/3} \Lambda_1}{3 \left( \Lambda_2 + \sqrt{-4 \Lambda_1^3 + \Lambda_2^2} \right)^{1/3}}\ +\  
\frac{ \left( \Lambda_2 + \sqrt{-4 \Lambda_1^3 + \Lambda_2^2} \right)^{1/3} }{3 \cdot 2^{1/3}}
\end{equation}
\smallskip

Here, \(\Lambda_1\) = 12\(P_0\) + \(P_2^2\) - 3\(P_1 P_3\) and \(\Lambda_2\) = 27\(P_1^2\) - 72\(P_0 P_2\) + 2\(P_2^3\) - 9\(P_1 P_2 P_3\) + 27\(P_0 P_3^2\). From equation (12) and (13), the solutions in \(\Omega\) might be positive, negative, or complex in nature. This purely depends on the magnitudes and signs of the parameters involved. When the continuous wave (cw) state is stable and real-valued, MI is not observed. The instability arises when the cw state is an imaginary value, and the growth rate of the instability in the MI is defined by gain, which is represented as follows:
\begin{equation}
\xi = \left\{ |\operatorname{Im}(\Omega)| \right\}_{\max}
\end{equation}
{The eigenfrequencies are obtained from the analytical solutions Eq. (14) and Eq. (15). Instability occurs when at least one eigenfrequency acquires a nonzero imaginary part. Here, the modulation instability gain $\xi$ is defined as the maximum imaginary part of the excitation frequency taken over all eigen-branches.} These solutions obtained in Equations~(12) and (13) would be used to understand the different physical effects that would be observed by varying the various parameters governing our system. The dispersion relation and, inevitably, the gain spectra would be significantly influenced by the Helicoidal SOC and Rabi Coupling. Furthermore, the nature of the atomic interaction strength plays a crucial role in the creation and sustenance of the MI. For mathematical convenience, we have assumed the density of the two components equal to 1, implying that $n_{10} = n_{20} = 1$. Further ahead, we have analyzed the influence of various parameters on the instability spectrum.

\section{Analysis of the Modulation Instability}
In this section, we analyze the impact of atomic interactions on the MI followed by the impact of helicoidal SOC and Rabi coupling on the MI.
\subsection{Impact of atomic-interactions on the MI}
Here we analyze the impact of atomic interactions by varying both the intra-species and inter-species interaction terms by keeping some of the other determining parameters at fixed values. 
{All parameters in this work are expressed in dimensionless units scaled by the characteristic harmonic trap frequency. 
Figure 1 displays the plot of gain with respect to the inter-species interaction term with different sets of values of Helicoidal SO coupling strength and Rabi coupling strength terms. Here, the other parameters have been fixed at, \( g_1 = 1 \), \( g_2 = 1 \), \( n_{01} = 1 \), \( n_{02} = 1 \), \( k_x = 1 \), \( k_y = 1 \) and \( \beta = 0.5 \). In order to distinguish the variation clearly, both \( \alpha \) and \( R \) have been varied simultaneously. 
Within this scaling, the Rabi coupling strength $R$ and the SOC strength $\alpha$ are independent control parameters. Values such as $R=0.5,1.5,3$ correspond to regimes where the coupling strength is comparable to or moderately larger than the trap frequency, which is experimentally accessible. Larger values are included to explore the strong-coupling regime and to illustrate qualitative trends in the modulation instability. The choice $\alpha = R$ is adopted to highlight the combined influence of spin–orbit and Rabi couplings.}
The helicoidal SOC strength \( \alpha \) and Rabi-coupling strength (R) are clearly dependent on the gain profile. In particular, greater values of \( \alpha \) and R initially lead to the increased MI gain, indicating that stronger inter-component and spin-orbit coupling amplify the instability in the early phases. In contrast to the situation with lower \( \alpha \) and R, the MI gain for these stronger couplings also decays more rapidly. Thus, indicating a more severe suppression of instability beyond a point. Moreover, as the inter-species contact gets closer to the regime of repulsion, the gain almost disappears. This trend demonstrates how attractive interactions between species can promote instability and the shift from repulsion to a more stable condensate state.
\begin{figure}[h]
    \centering
    \includegraphics[width=0.515\textwidth]{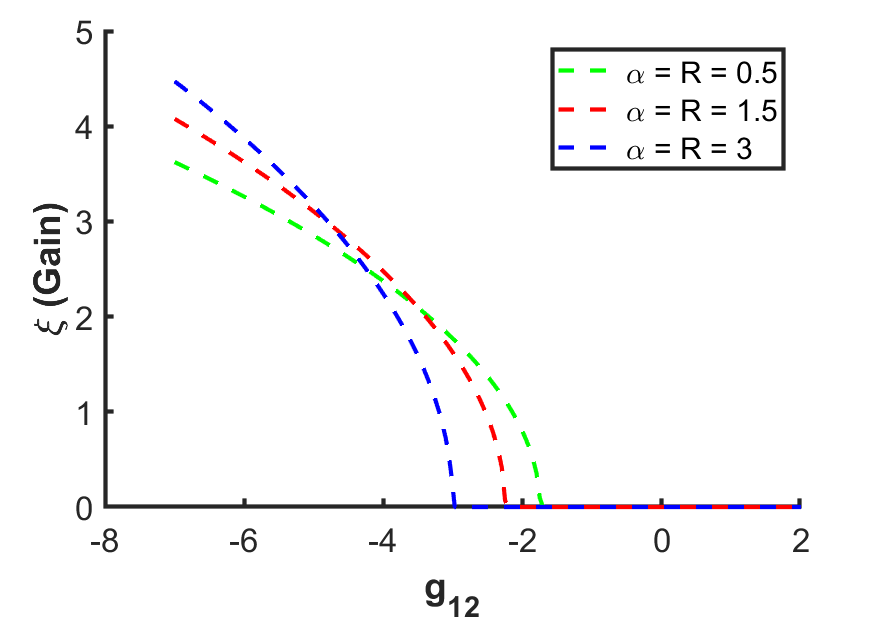}
    \caption{The MI gain \( \xi \) plotted against the inter-species parameter \( g_{12} \) for different Helicoidal SOC strength \( \alpha \) and Rabi coupling strength R.}
\end{figure}

Further, in Figure 2 we see the plot of MI gain with respect to \( k_x \), the wavenumber along the x-direction. This has been plotted for different sets of values of the intra-species parameters \( g_1 \) and \( g_2 \). Rest of the parameters have been fixed at, \( g_{12} = 2 \), \( n_{01} = 1 \), \( n_{02} = 1 \), \( k_y = 1 \), \( \beta = 0.5 \), \( \alpha = 0.5 \) and \( R = 0.5 \). It's observed that, when the intra-species interaction parameters \( g_1 \) and \( g_2 \) are raised from 0.2 to 0.8, the MI gain shows a nearly linear decrease in the peak amplitude. Hence, pointing to the stabilizing effect arising due to the increased repulsion within each component of the condensate. It's interesting to note that this linear trend is broken at \( g_1 \) = \( g_2 \) =1.0, suggesting a potential threshold beyond which the system's response experiences alterations. Furthermore, when \( g_1 \) and \( g_2 \) increase, the instability region's bandwidth gradually narrows, indicating lower susceptibility to density variations. Overall, the findings support the idea that strengthening the condensate's resistance to disturbances while boosting intra-species repulsion, reduces MI. {It should be emphasized that the modulation instability gain represents the exponential growth rate of unstable modes and does not correspond to a conserved spectral quantity; therefore, no momentum-space sum rule is expected, particularly in the presence of spin–orbit coupling and Rabi-induced coherence.}
\begin{figure}[h]
    \centering
    \includegraphics[width=0.48\textwidth]{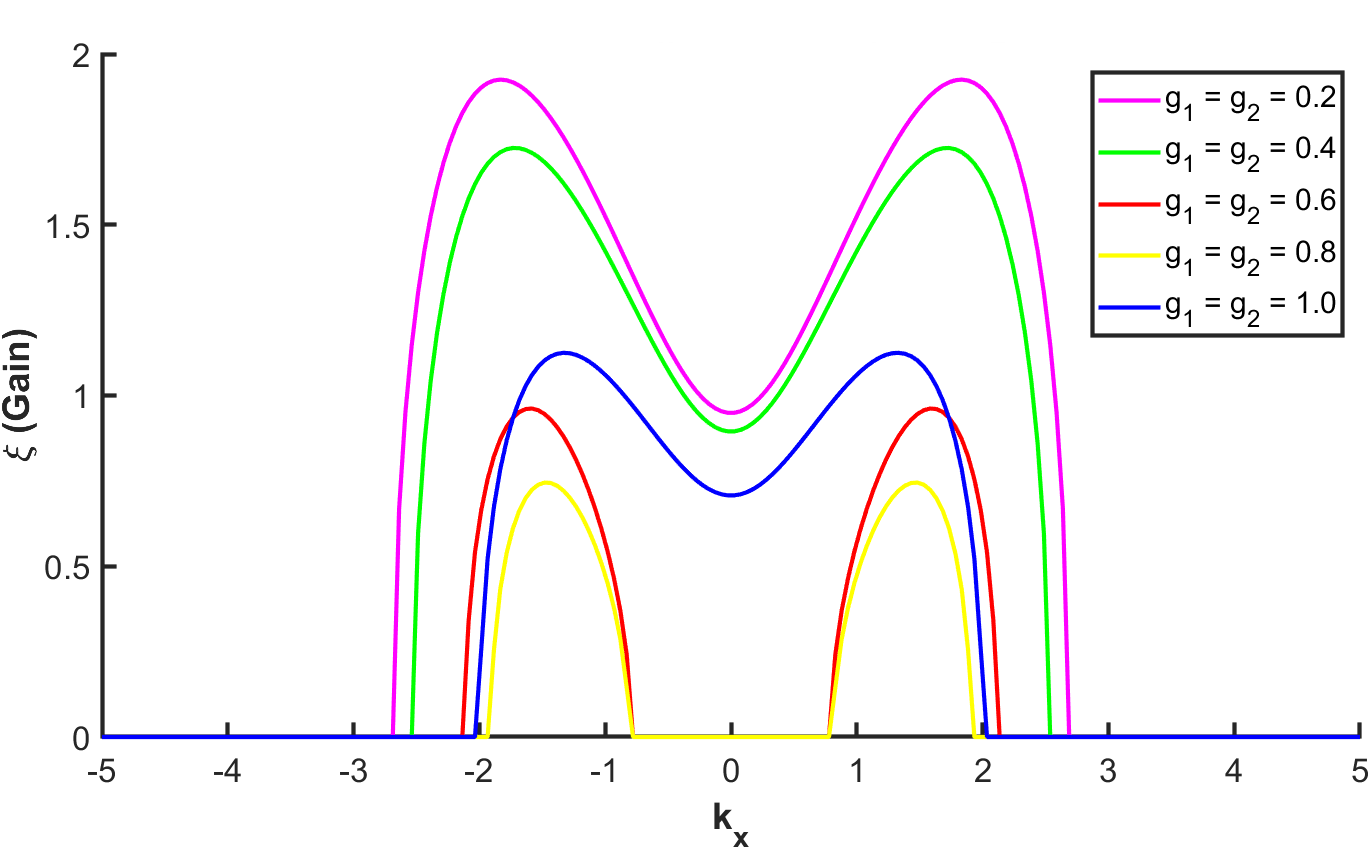}
    \caption{The MI gain \( \xi \) plotted against \( k_x \) for different intra-species interaction parameters \( g_1 \) and \( g_2 \).}
\end{figure}

In Figure 3, the MI gain has been obtained with respect to the wavenumbers \( k_x \) and \( k_y \), with plots (b), (d) and (f) representing the top view of the plots (a), (c) and (e) respectively. Except for the term of inter-species interaction \( g_{12} \), the rest of the parameters have been kept constant as follows; $g_1 = 1$, $g_2 = 1$, $n_{01} = 1$, $n_{02} = 1$, $\beta = 1$, $\alpha = 0.5$ and $R = 0.5$. For plots (a) and (b), \( g_{12} = 2 \); for plots (c) and (d) \(g_{12} = 3 \); and for plots (e) and (f) the value of \(g_{12}=4\). With the gradual increase of the inter-component interaction strength \(g_{12}\) into the repulsive regime, the peak of the hollow cylindrical-shaped modulation instability (MI) gain also increases. It should be noted that the instability region remains symmetric in the momentum space about \(k_x\) and \(k_y\). In addition, the top-view plots show the broadening of the red-ring structure with increasing \(g_{12}\), indicating that the radius of the unstable momentum-space ring expands along both directions. This implies a wider instability bandwidth in k-space, where MI becomes stronger and more active over a broader range of momentum states. In conclusion, a larger instability zone in the momentum space and a higher MI gain peak are the results of increasing \(g_{12}\).
\begin{figure}[h]
\centering
\xincludegraphics[width=0.475\linewidth,label=\textbf{a.}]{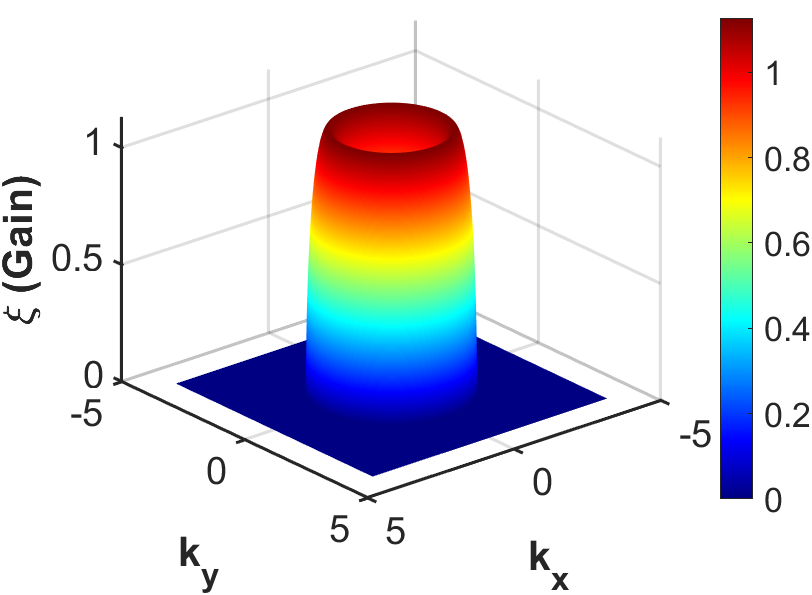}$\quad$
\xincludegraphics[width=0.475\linewidth,label=\textbf{b.}]{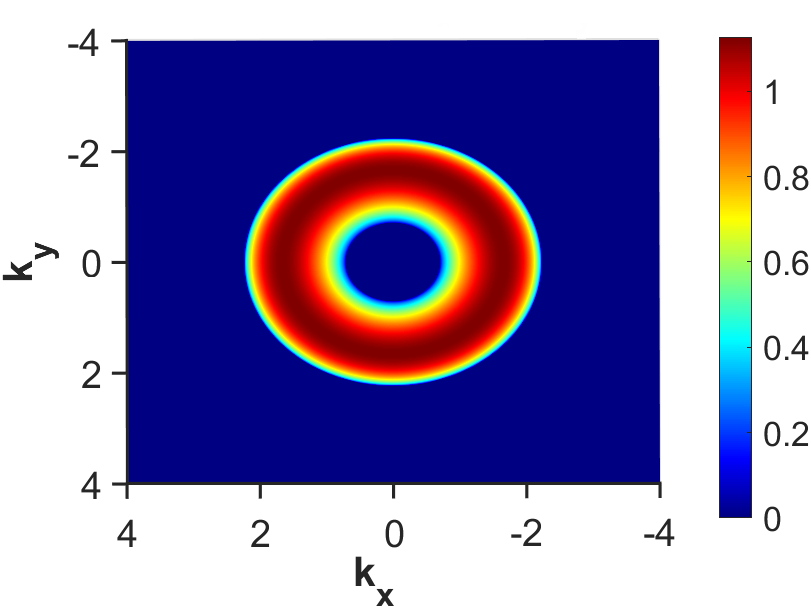}
\xincludegraphics[width=0.475\linewidth,label=\textbf{c.}]{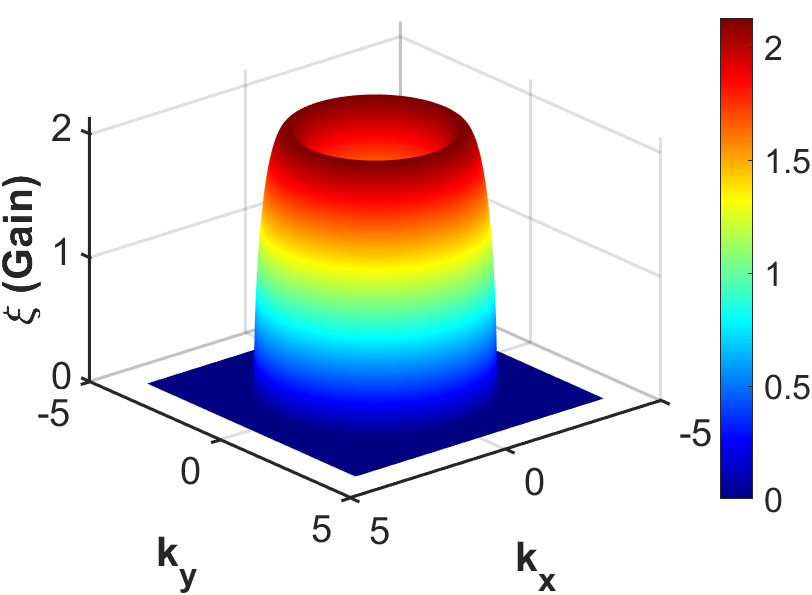}$\quad$
\xincludegraphics[width=0.475\linewidth,label=\textbf{d.}]{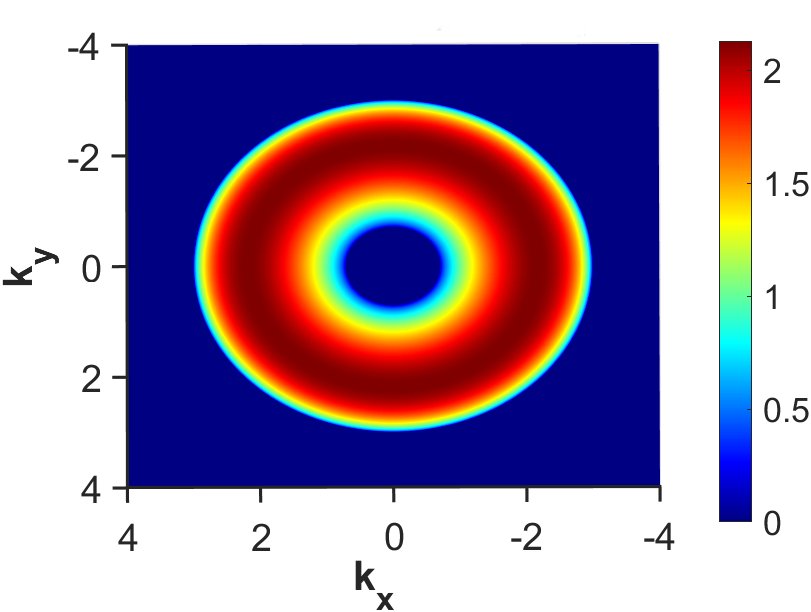}
\xincludegraphics[width=0.475\linewidth,label=\textbf{e.}]{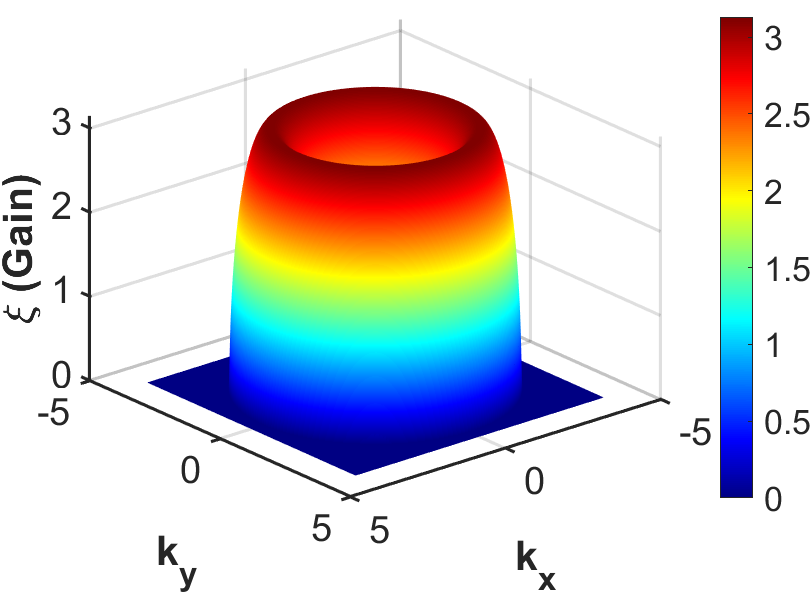}$\quad$
\xincludegraphics[width=0.475\linewidth,label=\textbf{f.}]{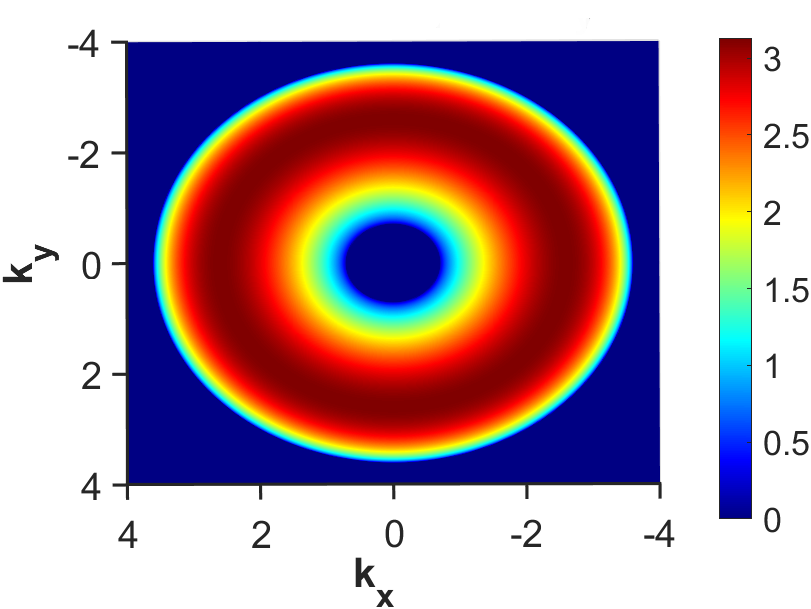}
\caption{Three dimensional (3D) surface plot of MI gain \( \xi \) presented with respect to \( k_x \) and \( k_y \) as depicted in plots (a), (c) and (e) with its corresponding two-dimensional (2D) contour plots depicted in (b), (d) and (f) respectively. Here, only \( g_{12}\) increases as we move from top to bottom with other values fixed for all plots as : $g_1 = 1$, $g_2 = 1$, $n_{01} = 1$, $n_{02} = 1$, $\beta = 1$, $\alpha = 0.5$ and $R = 0.5$.\ For plots (a) and (b), \( g_{12} = 2 \); for plots (c) and (d) \(g_{12} = 3 \); and for plots (e) and (f) the value of \(g_{12}=4\). }
\end{figure}

Figure 4 demonstrates the modulation instability (MI) gain which has been represented as a function of the momentum component \( k_x \) and the inter-species interaction parameter \( g_{12} \). Here, plots (a)-(b) correspond to excitation branches \( \Omega_{1,2} \), and (c)-(d) to excitation branches \( \Omega_{3,4} \). Plot Panels (a) and (c) depict the 2D contour plots, while (b) and (d) represent the corresponding 3D surface plots. A key finding is the localization of MI gain at larger \( |k_x| \) which indicates that modes with higher momenta do experience stronger amplification. This reflects the generation of spatially modulated instabilities, an authentic signature of complex nonlinear behavior in the system. Additionally, it can be observed that the MI gain increases with higher values of \( g_{12} \), indicating that stronger inter-species coupling, despite its repulsive or attractive nature, does enhance instability. This trend is evident across both the excitation branches. The gain profiles also exhibit symmetry around zero momenta along the x-direction. The fixed Rabi coupling preserves inter-component coherence, collectively giving shape to a rich instability landscape valid for pattern formation and nonlinear excitations in spin-orbit-coupled BECs.

\begin{figure}[h]
\centering
\xincludegraphics[width=0.475\linewidth,label=\textbf{a.}]{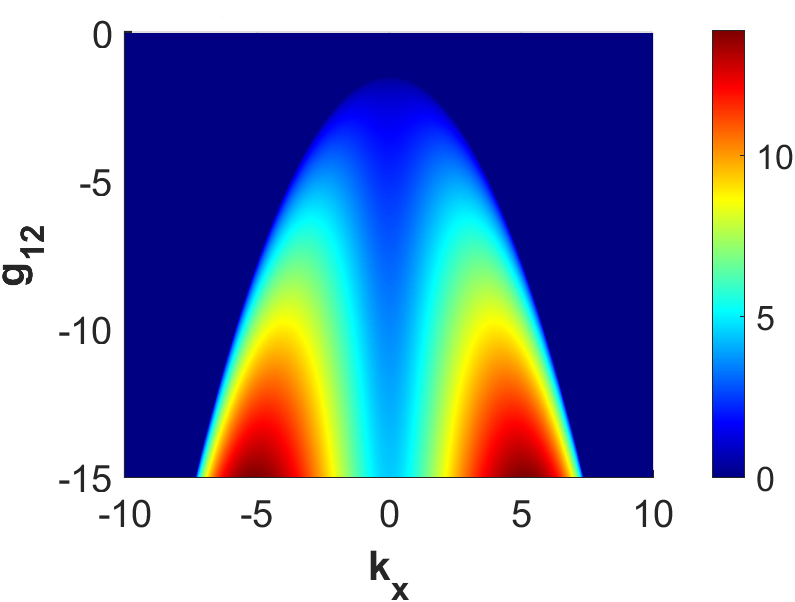}$\quad$
\xincludegraphics[width=0.475\linewidth,label=\textbf{b.}]{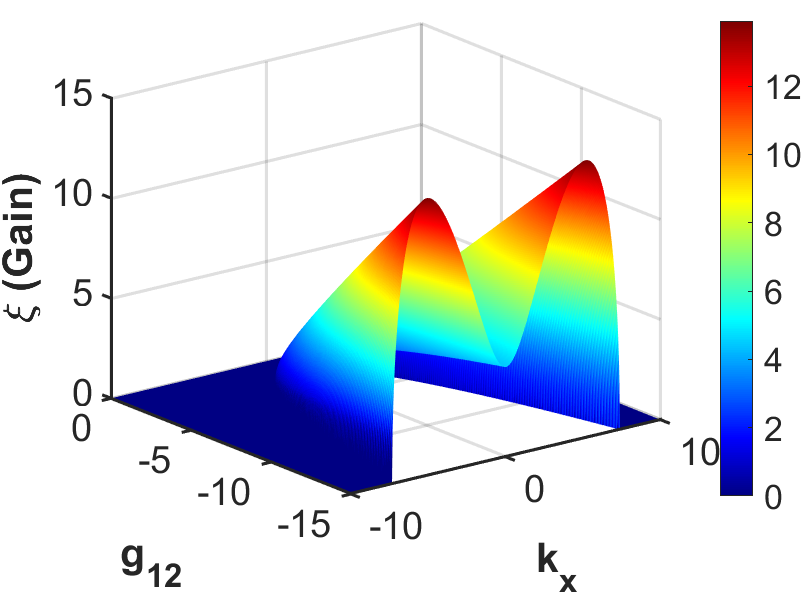}
\xincludegraphics[width=0.475\linewidth,label=\textbf{c.}]{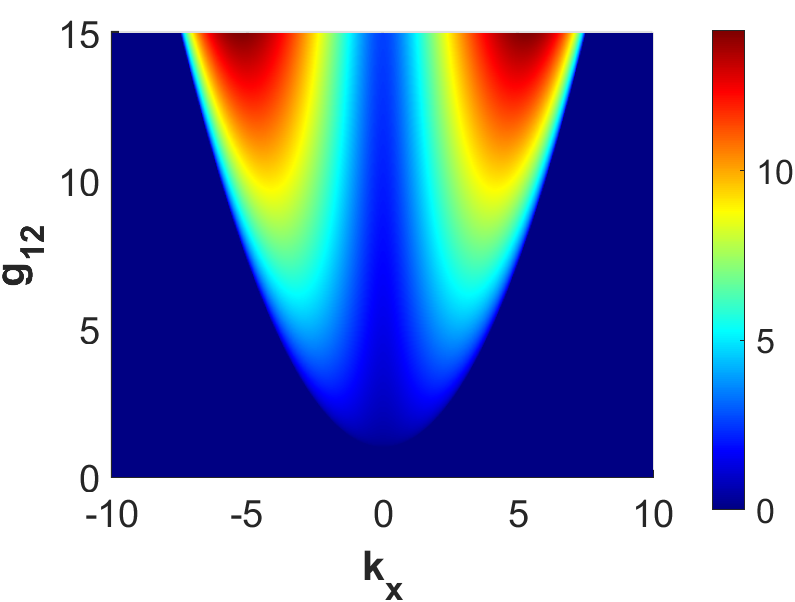}$\quad$
\xincludegraphics[width=0.475\linewidth,label=\textbf{d.}]{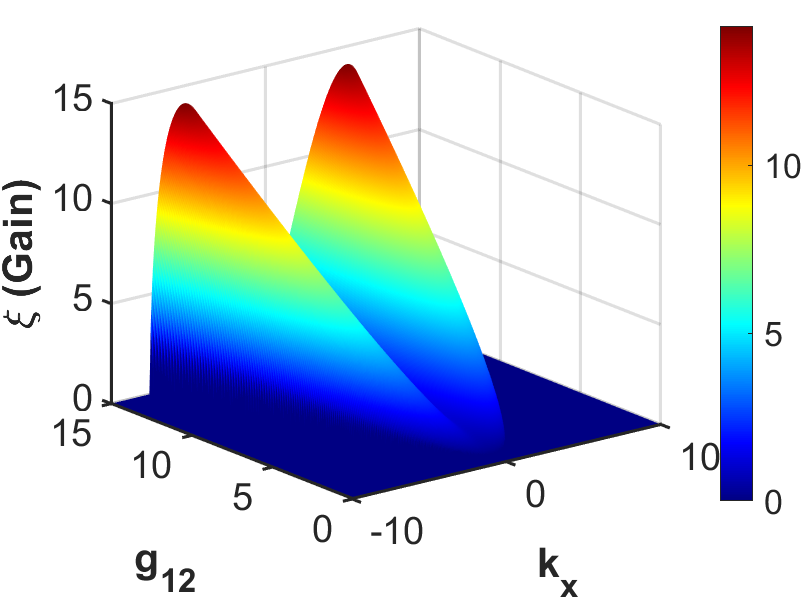}
\caption{Two dimensional (2D) contour panels of MI gain \( \xi \) plotted against \( k_x \) and inter-species interaction parameter \( g_{12} \), as depicted in plots (a) and (c) with its corresponding three dimensional (3D) surface plots depicted in (b) and (d) respectively. Here, plot (a)-(b) represents \(\Omega_{1,2}\) and plot (c)-(d) represents \(\Omega_{3,4}\). The rest of the parameters have been fixed as: $g_1 = 1$, $g_2 = 1$, $n_{01} = 1$, $n_{02} = 1$, $k_y=1$, $\beta = 1$, $\alpha = 0.5$ and $R = 0.5$.   }
\end{figure}

\subsection{Impact of Helicoidal SOC and Rabi-coupling on the MI}
Under this section, we have analyzed the impact of the helicoidal SOC strength and Rabi-coupling strength on the MI gain, by keeping the other governing parameters at specific values. In Figure 5, the MI gain has been presented as a function of \( k_x \), the wavenumber along the x-direction. Three distinct sets of Helicoidal SOC strength and Rabi Coupling strength parameters have been considered and arranged in the order of increasing magnitude with \( g_1 = 1 \), \( g_2 = 1 \), \( g_{12} = 2 \), \( n_{01} = 1 \), \( n_{02} = 1 \), \( k_y = 1 \), and \( \beta = 0.5 \). Higher values of \( \alpha\) and R  have consistently resulted in greater gain peaks, as observed thorough the investigation of the MI gain spectrum throughout the momentum axis \( k_x\) . This confirms that the likelihood for instability is largely amplified by the combined influence of strong spin-orbit and Rabi couplings. Also, stronger coupling enhances the formation of MI at higher momenta, as evidenced by the instability bands shifting toward larger \( |k_x|\) values, as \( \alpha\) and R grow. Notably, there is a drop at low \( |k_x|\) in the green curve that represents weak coupling circumstances  (\( \alpha\) = R = 0.5), suggesting that instability is suppressed at lower momenta. On the other hand, stronger coupling do not exhibit this dip, indicating a wider and longer-lasting instability spectrum.

\begin{figure}[h]
    \centering
    \includegraphics[width=0.475\textwidth]{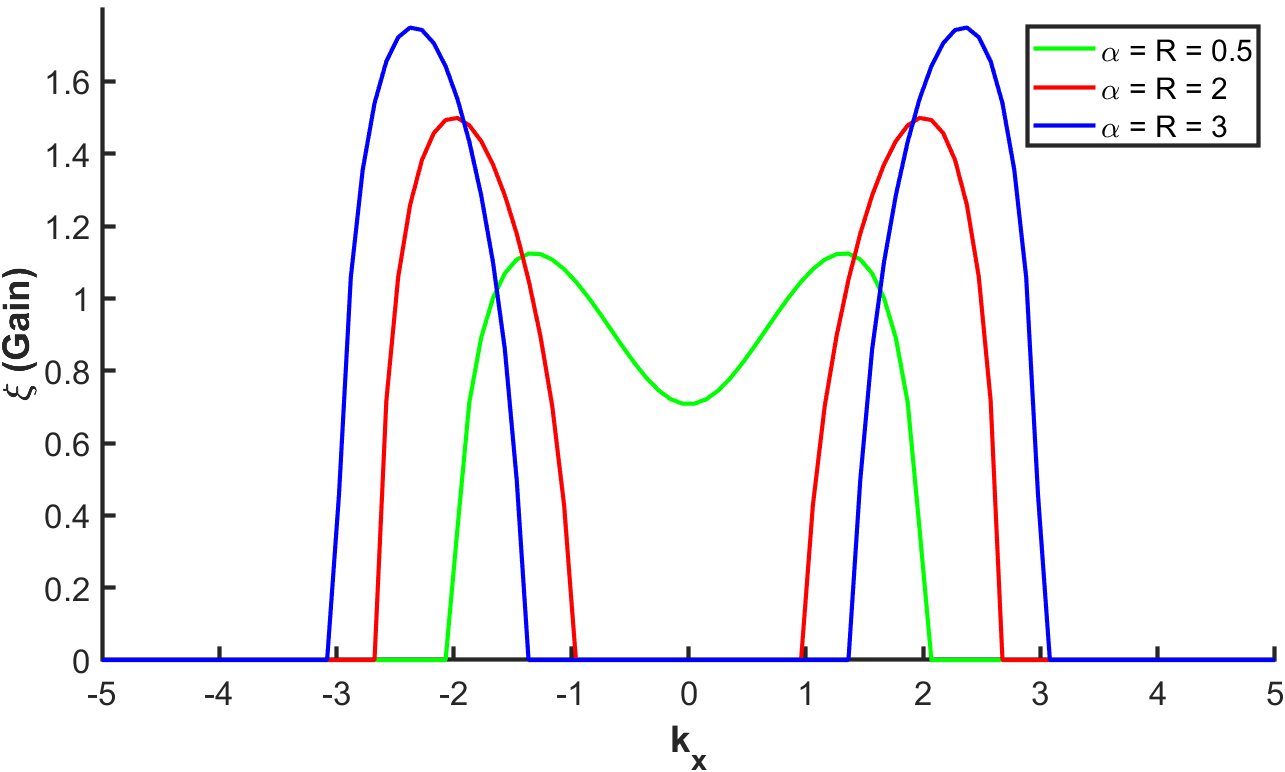}
    \caption{The MI gain \( \xi \) plotted against \( k_x \) for distinct values of Helicoidal SOC strength \(\alpha \) and Rabi coupling Strength R.}
\end{figure}

Furthermore, in Figure 6, we see MI gain plotted with respect to the Rabi-coupling strength parameter, considering different sets of values of the intra-species interaction terms. The remaining defining parameters have been fixed as, \( g_{12} = 2 \), \( n_{01} = 1 \), \( n_{02} = 1 \), \( k_x = 1 \), \( k_y = 1 \), \( \beta = 0.5 \), and \( \alpha = 0.5 \). The very first observation is the fact that for all the cases the MI gain is zero outside a specific width of R values, suggesting a well-defined instability region. As \( g_{1} = g_2\) increases from a value of 1 to 2, an increase in the MI gain is observed, affecting a slightly border range of R values as well. This indicates that stronger intra-species interaction affects instability growth yet a conclusive pattern can't be drawn due to the exception arising for intra-species interaction strength values of \( g_{1} = g_2 = 1.5\). Additionally, all the three curves appear symmetrical about R=0, indicating that MI is equally likely for both positive and negative values of R. Furthermore, the region around R=-1 to R=4 seems more susceptible for the growth of the instability.

\begin{figure}[h]
    \centering
    \includegraphics[width=0.475\textwidth]{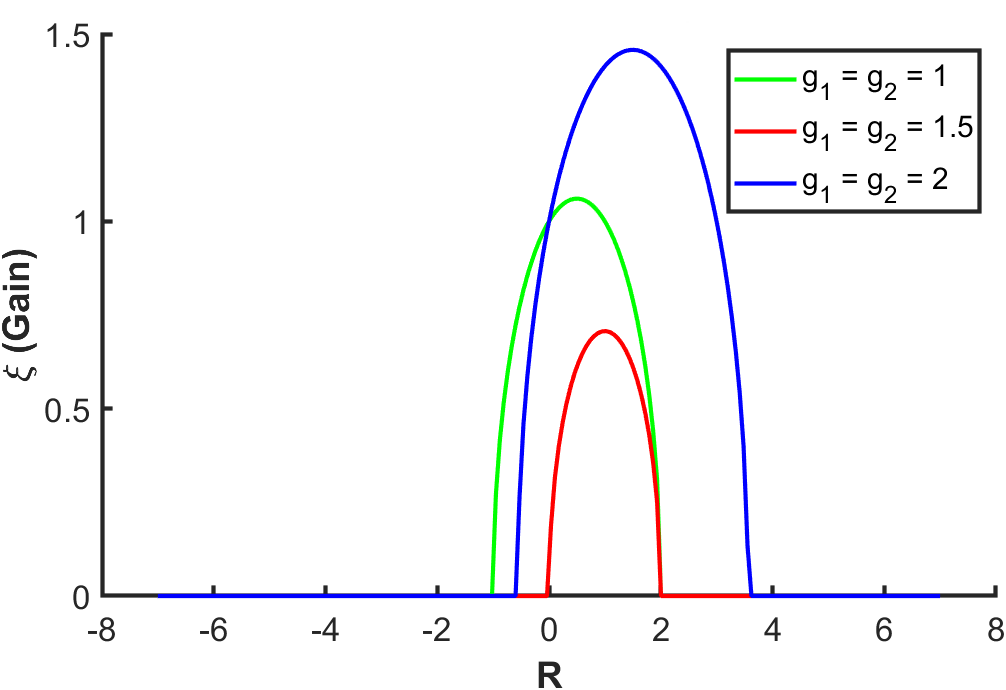}
    \caption{The MI gain \( \xi \) plotted against Rabi-coupling strength R, for varying intra-species interaction terms \(g_1\) and \( g_2\).}
\end{figure}

In Figure 7, the modulation instability (MI) gain \( \xi \) has been represented as a function of the wavenumbers \( k_x \) and \( k_y \). Here sub-figures (a), (c), and (e) display the three-dimensional (3D) surface plots, and (b), (d), and (f) show their corresponding two-dimensional (2D) top-view contour representations. For this investigation the helicoidal SOC strength \( \alpha \) and the Rabi-coupling term \( R \) are gradually increased from left to right, while all other parameters have been fixed as follows: \( g_1 = 1 \), \( g_2 = 1 \), \( g_{12} = 2 \), \( n_{01} = 1 \), \(n_{02} = 1 \), and \( \beta = 1 \). As \( \alpha \) and \( R \) increase across sub-figures, a significant enhancement in the peaks of the MI gain \( \xi \) is observed, indicating that stronger helicoidal SO interactions and Rabi-coupling intensify the MI in the system. Furthermore, the 3D plots reveal a protrusive ring-shaped MI structure in momentum space, which continues to maintain a cylindrical symmetry around the origin. This remains consistent with the system's isotropic configuration in the \( k_x \)-\( k_y \) plane. In addition, the 2D top view plots (b), (d) and (f) clearly show a progressive expansion in the radius of the instability ring. This signifies that the instability region shifts for a higher magnitude of momentum values, while the bandwidth of the momenta of particle that it affects remains more or less the same. Therefore, the combined influence of an increased magnitude of helicoidal SOC and Rabi coupling not only increases the magnitude of the instability but also extends the region in momentum space over which this instability persists.

\begin{figure}[h]
\centering
\xincludegraphics[width=0.475\linewidth,label=\textbf{a.}]{fig3a.png}$\quad$
\xincludegraphics[width=0.475\linewidth,label=\textbf{b.}]{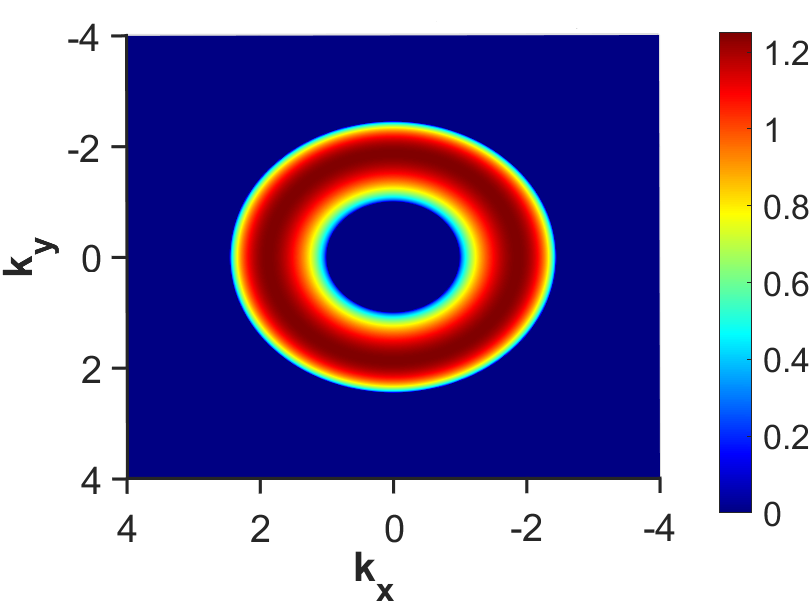}
\xincludegraphics[width=0.475\linewidth,label=\textbf{c.}]{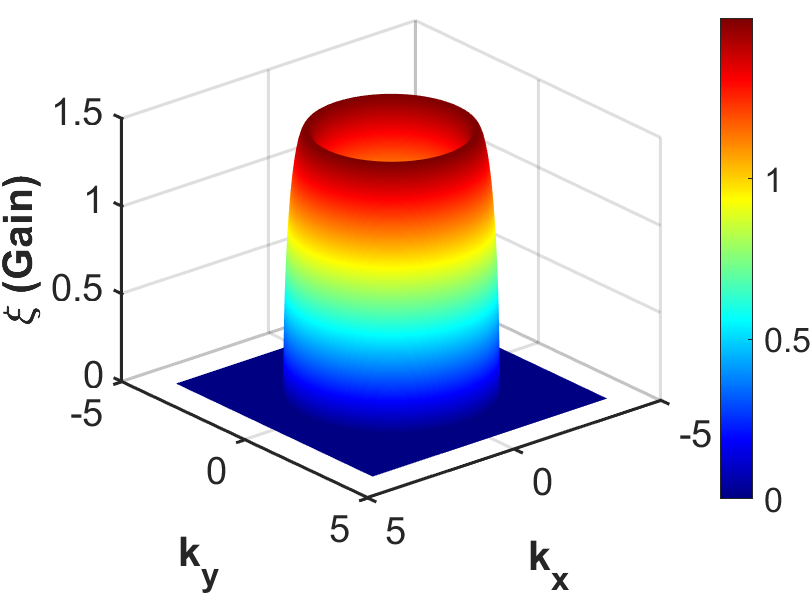}$\quad$
\xincludegraphics[width=0.475\linewidth,label=\textbf{d.}]{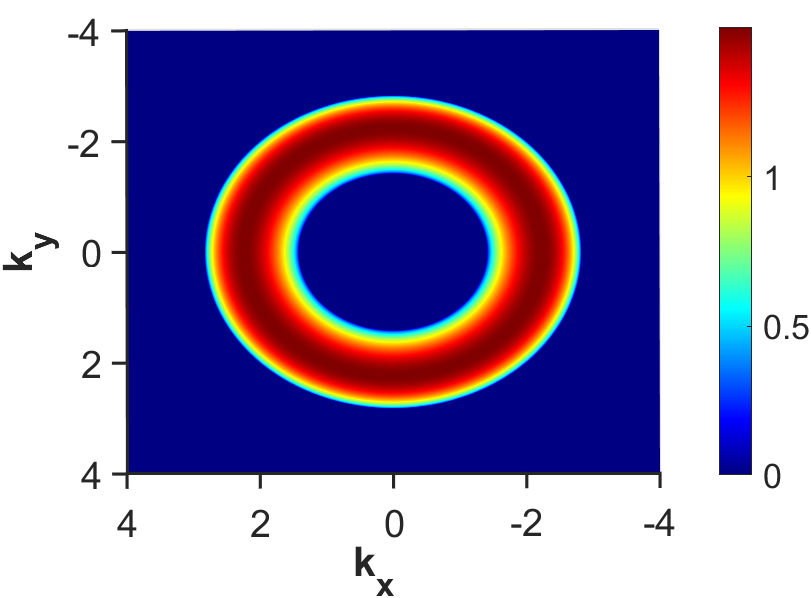}
\xincludegraphics[width=0.475\linewidth,label=\textbf{e.}]{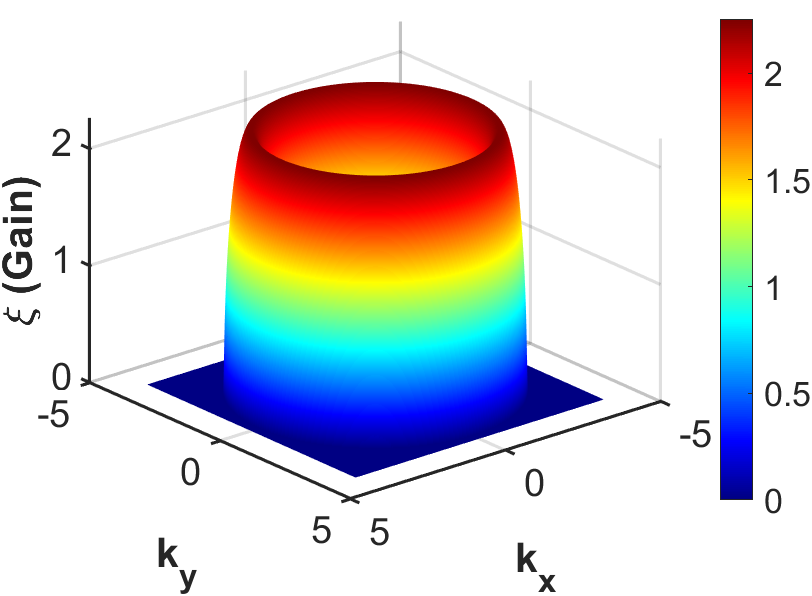}$\quad$
\xincludegraphics[width=0.475\linewidth,label=\textbf{f.}]{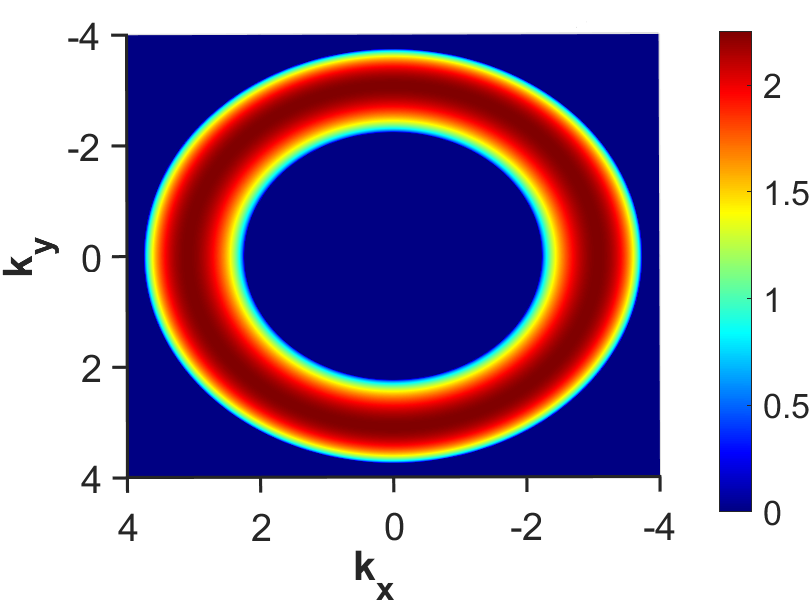}
\caption{Three-dimensional (3D) surface plots of the modulation instability (MI) gain $\xi$ as a function of the wave-vector components $k_x$ and $k_y$ are shown in panels (a), (c), and (e), with the corresponding two-dimensional (2D) contour plots presented in panels (b), (d), and (f), respectively. The spin--orbit coupling strength $\alpha$ and the Rabi coupling $R$ are increased simultaneously from top to bottom, while all other parameters are kept fixed at $g_1 = g_2 = 1$, $g_{12} = 2$, $n_{01} = n_{02} = 1$, and $\beta = 1$. Panels (a,b) correspond to $\alpha = R = 0.5$; panels (c,d) to $\alpha = R = 1$; and panels (e,f) to $\alpha = R = 1.5$.}
\end{figure}

\begin{figure}[h!]
\centering
\xincludegraphics[width=0.475\linewidth,label=\textbf{a.}]{fig3a.png}$\quad$
\xincludegraphics[width=0.475\linewidth,label=\textbf{b.}]{fig3b.png}
\xincludegraphics[width=0.475\linewidth,label=\textbf{c.}]{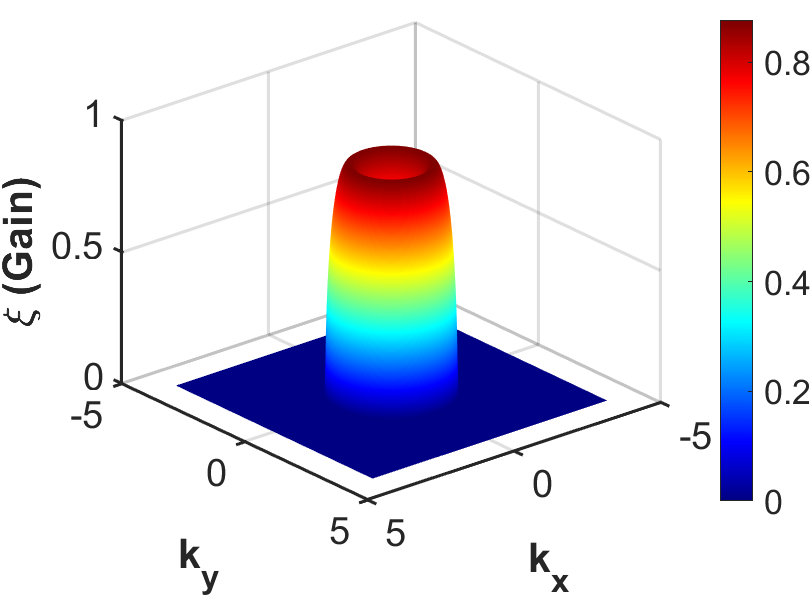}$\quad$
\xincludegraphics[width=0.475\linewidth,label=\textbf{d.}]{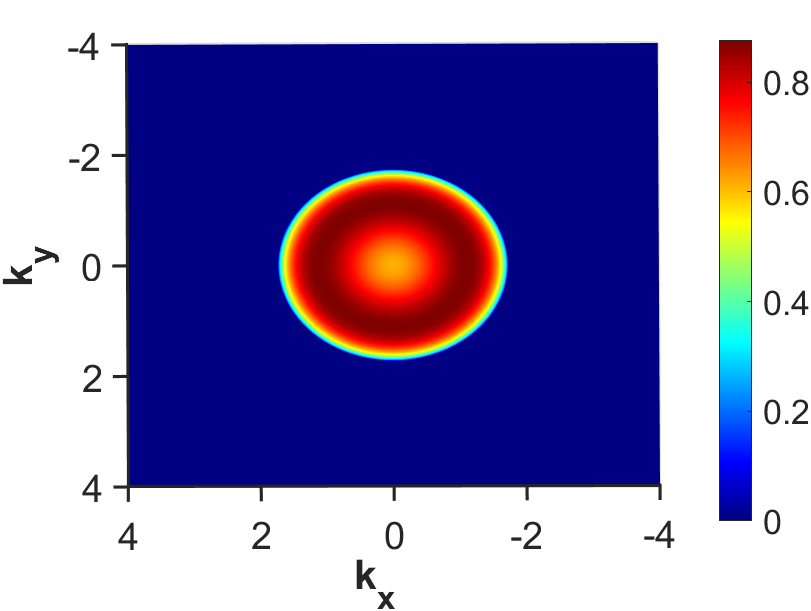}
\xincludegraphics[width=0.475\linewidth,label=\textbf{e.}]{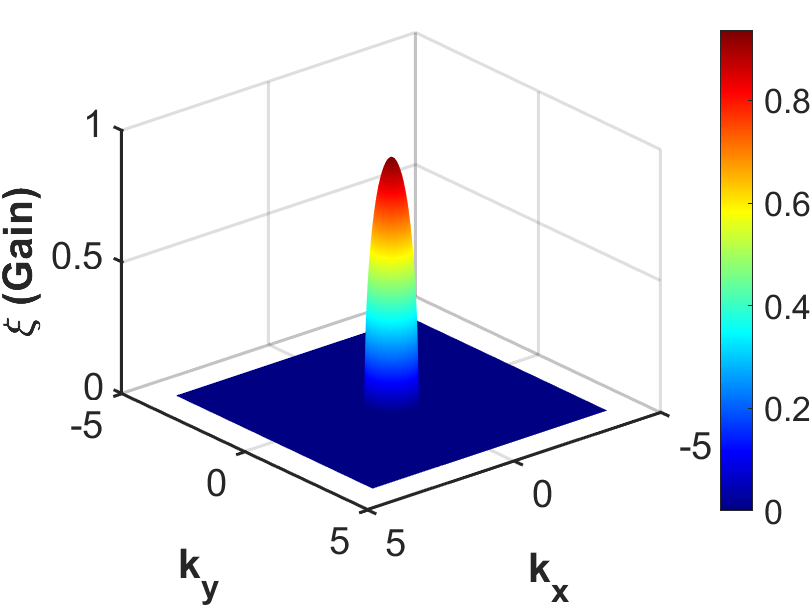}$\quad$
\xincludegraphics[width=0.475\linewidth,label=\textbf{f.}]{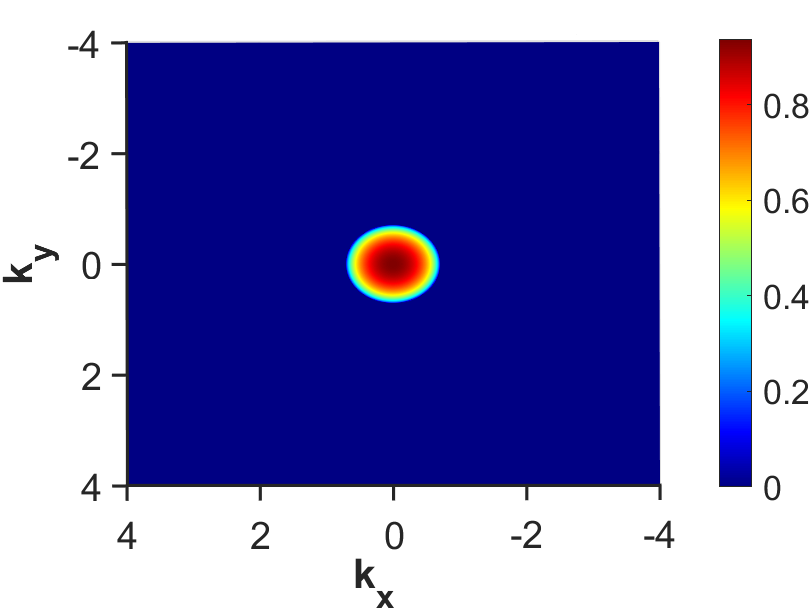}
\caption{{Three-dimensional (3D) surface plots of the modulation instability (MI) gain $\xi$ as a function of $k_x$ and $k_y$, shown in panels (a), (c), and (e), together with the corresponding two-dimensional (2D) contour plots in panels (b), (d), and (f), respectively. The Rabi coupling strength $R$ is varied from top to bottom, while all other parameters are kept fixed at $g_1 = g_2 = 1$, $g_{12} = 2$, $n_{01} = n_{02} = 1$, $\beta = 1$, and $\alpha = 0.5$. Panels (a)--(b) correspond to $R = 1$, panels (c)--(d) to $R = -0.5$, and panels (e)--(f) to $R = -1$.}}
\end{figure}
In Figure 8, the modulation instability (MI) gain \( \xi \) is investigated as a function of wavenumbers \( k_x \) and \( k_y \) along x and y directions respectively. The 3D surface plots have been shown in (a),(c), and (e) and their respective top-view 2D contour plots are shown in (b), (d) and (f). The main aim of this analysis is to assess the effect of the Rabi-coupling strength term \( R \) on the instability profile of our system when it is exposed to helicoidal spin-orbit coupling (SOC). The remaining parameters for this investigation are fixed as: \( g_1 = 1 \), \( g_2 = 1 \), \( g_{12} = 2 \), \( n_{01} = 1 \), \( n_{02} = 1 \), \( \beta = 1 \), and \( \alpha = 0.5 \). The parameter \( R \) is varied across the columns as follows: \( R = 1 \) for (a) and (b), \( R = -0.5 \) for (c) and (d), and \( R = -1 \) for (e) and (f). As the Rabi coupling strength is reduced from a positive to a negative value, a significant transformation in the MI profile is observed. In plot (a), having a positive Rabi coupling \( R = 1 \), the MI gain forms a pronounced ring-shaped structure in the momentum space as observed in the previously mentioned cases as well. The ring structure which is clearly seen in the top-view contour of plot (b), signifies a distributed instability across a wide range of momentum modes. The hollow core of this ring is centered at \( k_x = k_y = 0 \). Furthermore, as the Rabi coupling is reduced to a small negative value \( R = -0.5 \) in plots (c) and (d) it can be observed that the ring structure begins to shrink in radius and intensity, taking the shape of a compressed bullet. This is further accompanied by a narrowing of the MI bandwidth as well. This suggests a weakening of the MI as the instability reduces itself to lower mometa particles. This could potentially be due to the competing effects introduced by the negative Rabi term, which modifies the energy dispersion and suppresses long-range coherent exchange between the condensate components. Additionally, at \( R = -1 \) in plots (e) and (f), the MI region gets strongly localized into a narrow momentum space with a sharper bullet like central peak. The resulting MI gain is not only reduced in magnitude but the instability is confined to a small range near the origin of the \( k \)-space too. This behavior illustrates that strong negative Rabi coupling acts as a stabilizing factor, suppressing the modulational instability over almost the entire momentum domain. Thus, it can be said that increasing the magnitude of negative Rabi coupling \( R \) in the helicoidal SOC-BEC system could lead to a progressive suppression of MI, both in terms of amplitude and its spectral width.

\begin{figure}[h]
\centering
\xincludegraphics[width=0.485\linewidth,label=\textbf{a.}]{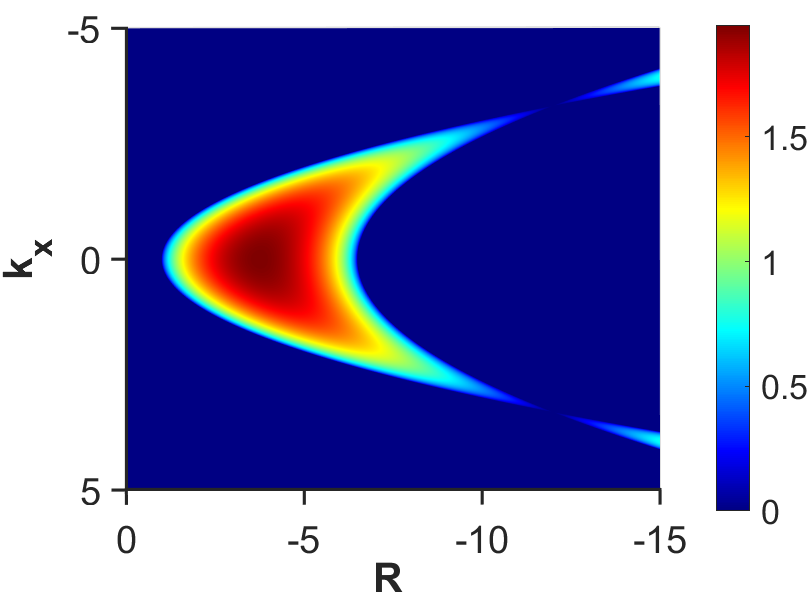}$\quad$
\xincludegraphics[width=0.46\linewidth,label=\textbf{b.}]{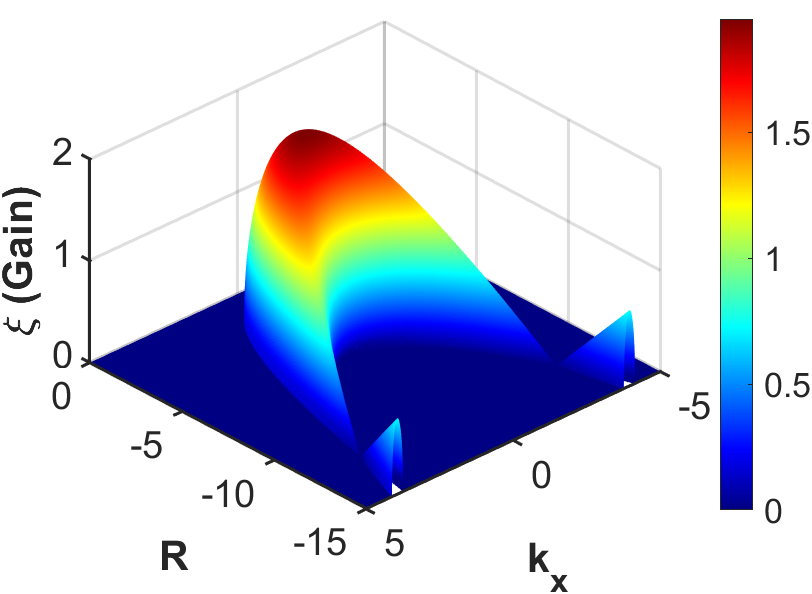}
\xincludegraphics[width=0.45\linewidth,label=\textbf{c.}]{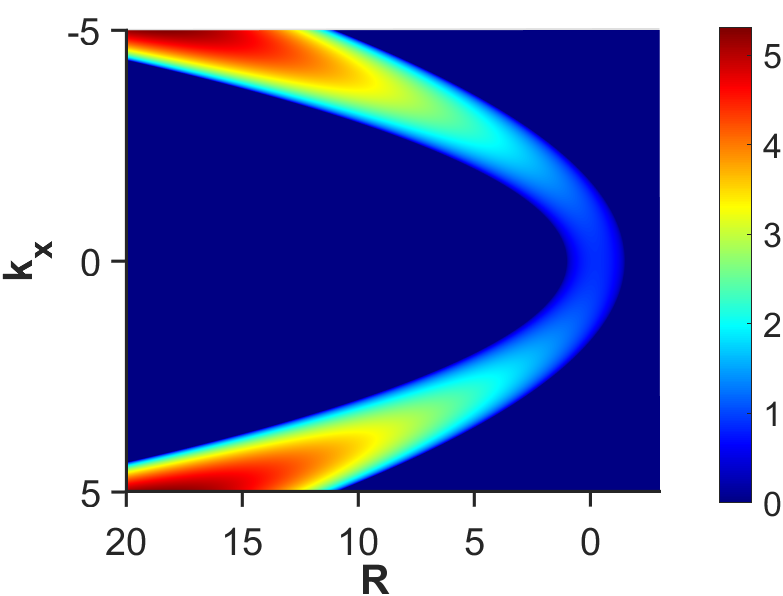}$\quad$
\xincludegraphics[width=0.45\linewidth,label=\textbf{d.}]{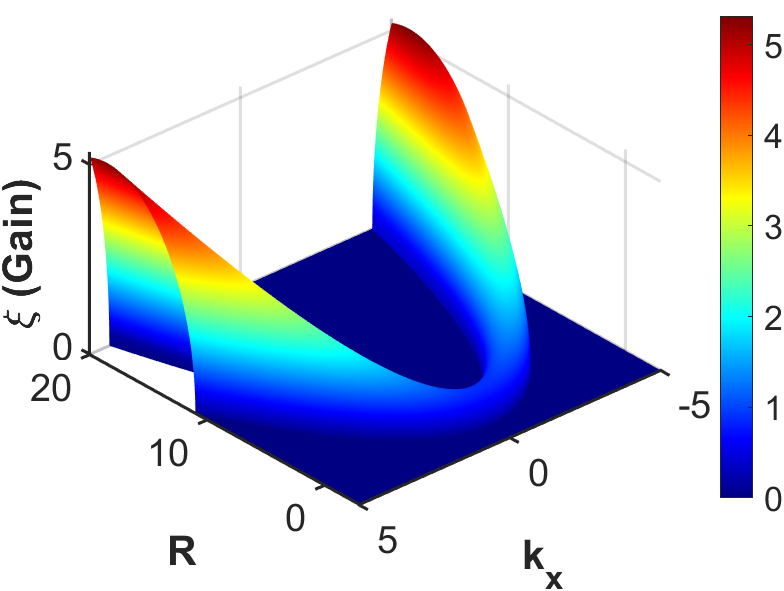}
\caption{Two dimensional (2D) contour panels of MI gain \( \xi \) plotted against \( k_x \) and Rabi coupling strength term R as depicted in plots (a) and (c) with its corresponding three dimensional (3D) surface plots depicted in (b) and (d) respectively. Here, plot (a)-(b) represents \(\Omega_{1,2}\) and plot (c)-(d) represents \(\Omega_{3,4}\). The rest of the parameters have been fixed for all plots as  $g_1 = 1$, $g_2 = 1$, $g_{12} = 1$ $n_{01} = 1$, $n_{02} = 1$, $k_y=1$, $\beta = 1$ and $\alpha = 1$}
\end{figure}

Figure 9 represents the MI gain \( \xi \) with respect to \( k_x \) and the Rabi coupling strength \( R \). Plots (a) and (c) show the 2D contours, while plots (b) and (d) provide the corresponding 3D surface plots for the same. The remaining parameters for all other plots are fixed as follows: \( g_1 = 1 \), \( g_2 = 1 \), \( g_{12} = 1 \), \( n_{01} = 1 \), \( n_{02} = 1 \), \( k_y = 1 \), \( \beta = 1 \), and \( \alpha = 1 \). In Fig.9(a), the gain is evidently localized around lower values of \( |k_x| \) and negative values of \( R \). This suggests that low-momentum modes experience enhanced amplification under moderately negative Rabi coupling, indicating enhanced long-wavelength instabilities which are influenced by the HSOC and R-induced spin coherence in the system. In contrast, as seen in Fig.9(c), the gain shifting toward higher \( |k_x| \) and more positive values of \( R \) indicates dominant short-wavelength mode amplification. This behavior has been confirmed in the 3D plots (b) and (d), further highlighting how increasing \( R \) promotes stronger MI at higher momenta. This crossover from long- to short-wavelength MI accompanied with varying \( R \) undermines its tunability too. The directional modulation brought about by the HSOC is further reflected in the anisotropic gain distribution, highlighting its crucial role in determining MI behavior.

\subsection{Impact of Both Atomic Interactions and Rabi Coupling on the MI}
In this subsection, we dive deeper into the combined influence of inter-species interaction strength \( g_{12} \) and Rabi coupling strength \( R \) on the modulation instability (MI) dynamics of this SO-coupled BEC under a two-dimensional harmonic potential. Unlike previous sections where either atomic interactions or spin-related couplings were varied independently, this analysis investigates the interplay between these two key parameters. To achieve this, we simultaneously tune both \( g_{12} \) and \( R \), in order to understand how this coupling between atomic and spin dynamics governs the localization and amplitude of the MI gain \( \xi \). This approach provides an in-depth understanding of how the competing contributions of nonlinear atomic interactions and coherent spin mixing jointly coordinate the growth of perturbations in the system. Figure 10 represents the modulation instability (MI) gain \( \xi \) plotted against the inter-species interaction parameter \( g_{12} \) and the Rabi coupling strength \( R \) for our system. Plots (a) and (c) show the 2D contour maps, while (b) and (d) show the corresponding 3D surface plots for the same. Rest of the parameters have been fixed for all plots as follows: \( g_1 = 1 \), \( g_2 = 1 \), \( n_{01} = 1 \), \( n_{02} = 1 \), \( k_x = 1 \), \( k_y = 1 \), \( \beta = 0.5 \), and \( \alpha = 0.5 \). In Fig.10(a), the MI gain is predominantly localized at larger \( R \) values and higher magnitudes of \( g_{12} \). This showcases that strong Rabi coupling and inter-species interactions jointly enhance the instability growth. Additionally we see that as \( g_{12} \) tends toward zero, the gain diminishes, reflecting the crcucial role of inter-component coupling in driving the MI. This trend is consistently observed in the surface plots (b) and (d) as well, where gain intensifies at the edges of the parameter space. These results emphasize that the interplay between the tunable Rabi coupling and atomic interaction governs the modulation instability landscape in similar systems.

\begin{figure}[h]
\centering
\xincludegraphics[width=0.475\linewidth,label=\textbf{a.}]{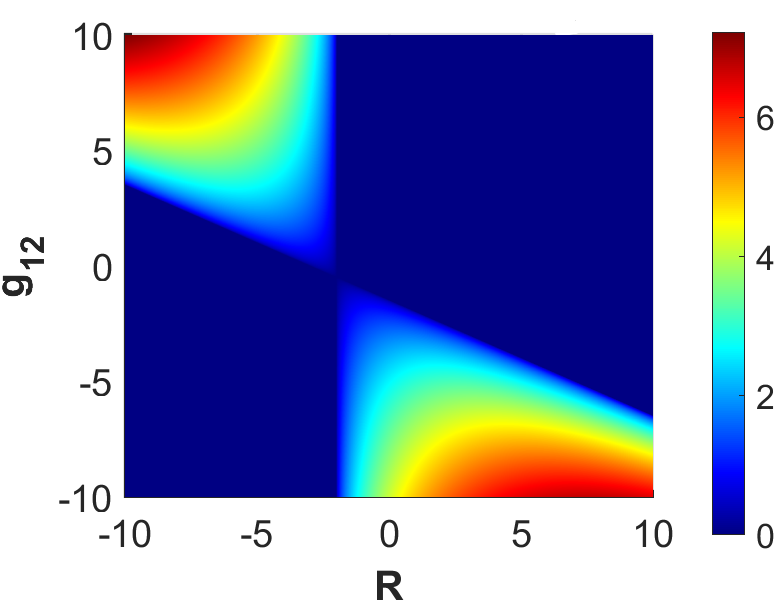}$\quad$
\xincludegraphics[width=0.475\linewidth,label=\textbf{b.}]{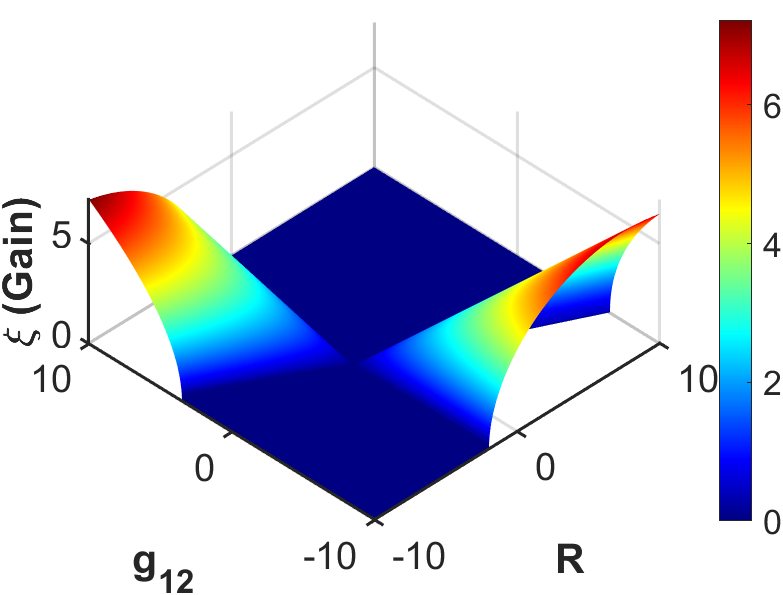}
\xincludegraphics[width=0.475\linewidth,label=\textbf{c.}]{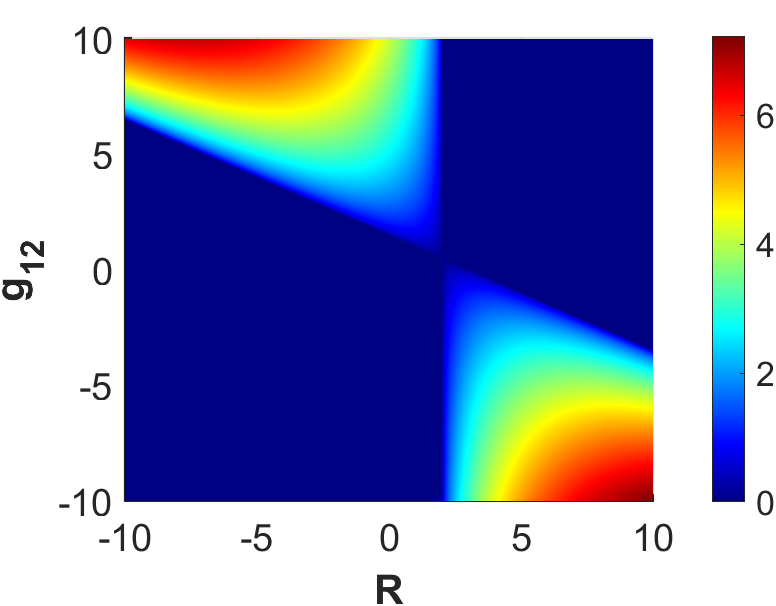}$\quad$
\xincludegraphics[width=0.475\linewidth,label=\textbf{d.}]{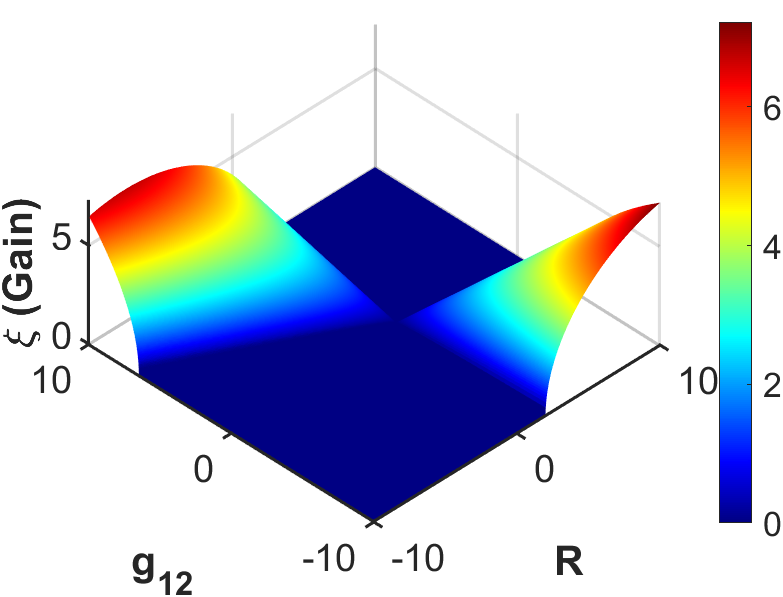}
\caption{Two dimensional (2D) contour panels of MI gain \( \xi \) plotted against \( g_{12} \) and Rabi-coupling strength term R as depicted in plots (a) and (c) with its corresponding three dimensional (3D) surface plots depicted in (b) and (d) respectively. Here, plot (a)-(b) represents \(\Omega_{1,2}\) and plot (c)-(d) represents \(\Omega_{3,4}\). The rest of the parameters have been fixed as $g_1 = 1$, $g_2 = 1$, $n_{01} = 1$, $n_{02} = 1$, $k_x=1$, $k_y=1$, $\beta = 0.5$ and $\alpha = 0.5$ . \  }
\end{figure}

\subsection{Perturbed Solution}

\begin{figure*}
    \centering \includegraphics[width=1.0\textwidth]{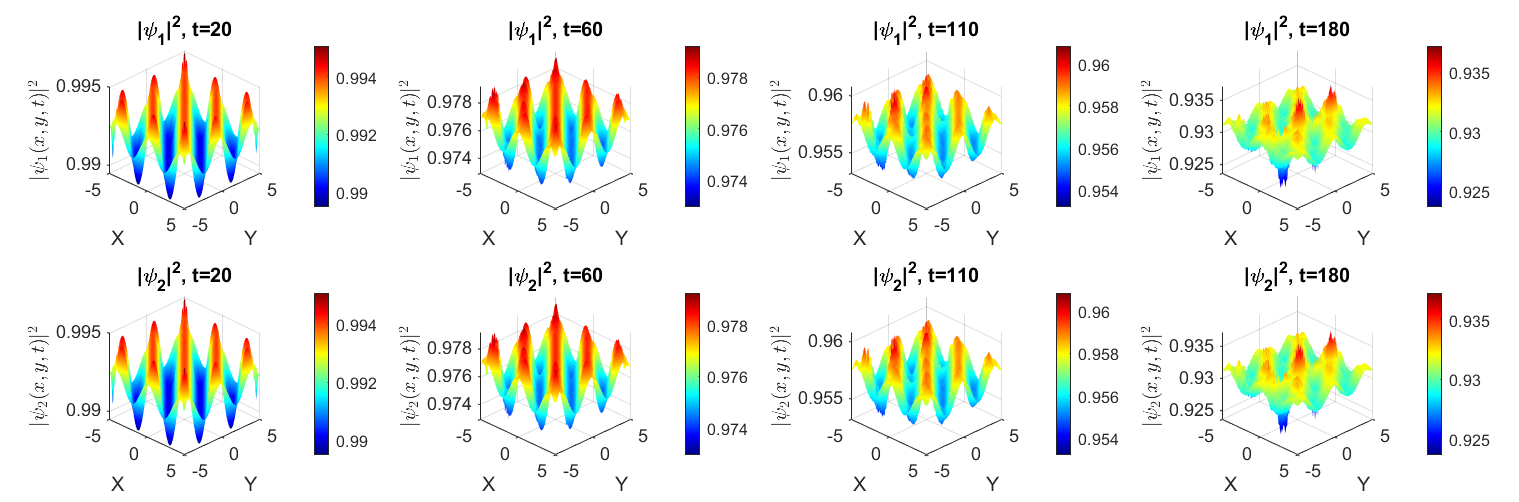}
    \caption{MI evolution in both condensates with increasing time in the presence of helicoidal spin--orbit coupling (HSOC) and Rabi coupling. Shown are density snapshots of $|\psi_{1}(x,y,t)|^{2}$ (top row) and $|\psi_{2}(x,y,t)|^{2}$ (bottom row) at times $t=20,\,60,\,110,\,180$. The system parameters are as follows: $\alpha=0.5$, $\beta=0.25$, Rabi coupling $\Omega=0.2$, intra-species interactions $g_{11}=g_{22}=0.5$, inter-species interaction $g_{12}=0.6$, and initial perturbation wave numbers $K_{x}=K_{y}=1.5$.}
\end{figure*}

The onset of modification instability (MI) is predicted by the linear stability analysis previously presented, but the long-term nonlinear evolution is not covered.  In order to solve this, we use the split-step Fourier method to directly numerically simulate the coupled Gross--Pitaevskii equations with helicoidal spin--orbit coupling (SOC) and Rabi interaction. The initial conditions are taken as weakly perturbed plane waves,
\begin{eqnarray}
\psi_{1}(x,y,0)=\psi_{2}(x,y,0)=\sqrt{n_{0}} +\epsilon \cos(K_{x}x+K_{y}y)\ ,
\end{eqnarray}

with $n_{0}=1$, $\epsilon=0.001$, and perturbation wavenumbers $K_{x}=K_{y}=1.5$. The parameters of the system are fixed as $\alpha=0.5$, $\beta=0.25$, $g_{11}=g_{22}=0.5$, $g_{12}=0.6$, and $\Omega=0.2$. 
The simulated time evolution of the densities $|\psi_{1}|^{2}$ and $|\psi_{2}|^{2}$ is displayed in Fig.~11 for representative snapshots at $t=20,\,60,\,110,$ and $180$.At first, both parts exhibit periodic modulations that are well-ordered and inherited from the seed perturbation.  The onset of MI is indicated as the modulation increases in amplitude over time and irregularities start to show up along both spatial directions.  The helicoidal SOC under consideration introduces anisotropic growth that is controlled by the interaction of $\alpha$ and Rabi coupling term strength term $\Omega$. This in contrast to other SOC cases like the favors of Rashba--Dresselhaus SOC case \cite{ref45}, where the instability frequently displays directional preferences linked to the underlying lattice symmetry.  As a result, density patterns become distorted and fluctuate along some directions while remaining comparatively regular along others.

The two condensates' near-synchronous behavior makes the Rabi coupling's function especially clear.  The densities $|\psi_{1}|^{2}$ and $|\psi_{2}|^{2}$ largely evolve in tandem, despite the emergence of local fluctuations. This is indicative of the phase-locking imposed by the intercomponent coupling.  The instability causes the initial periodic order to break down at later times ($t=110$ and $t=180$), resulting in delocalized and fluctuating density structures.  These findings demonstrate that although MI is driven by both helicoidal and Rashba--Dresselhaus couplings, the coexistence of helicoidal SOC and Rabi coupling creates a dynamical scenario that is qualitatively different and in which strong inter-component synchronization and anisotropic instability growth coexist as well.

\section{Trapping Frequency and Condensate Configuration}

To further broaden our understanding of the dynamics of the condensate thus obtained in the presence of helicoidal SOC and Rabi-coupling, under a symmetric two-dimensional harmonic trap, we have numerically solve the coupled Gross-Pitaevskii equations using the split-step Fourier method. The initial wavefunctions \(\psi_1(x, y, 0)\) and \(\psi_2(x, y, 0)\) are chosen as Gaussian profiles with equal widths and amplitudes, and the system is evolved up to a final time \(T = 3.0\) using a time step of \(\Delta t = 0.01\). 

\subsection*{Case 1: Isotropic Trap Configuration}
\begin{figure}[h]
    \centering
    \xincludegraphics[width=1\linewidth,label=\textbf{a.}]{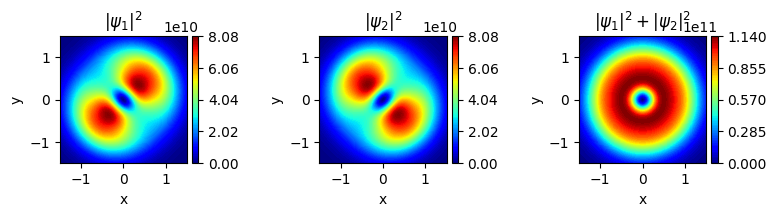}
    \xincludegraphics[width=1\linewidth,label=\textbf{b.}]{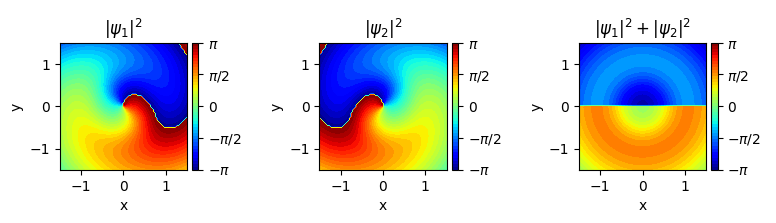}
    \caption{The spatial density distributions of the two-component wavefunctions \( \psi_1(x,y) \), \( \psi_2(x,y) \), and their total density \( |\psi_1|^2 + |\psi_2|^2 \) are displayed in the top panels (\textbf{a}). The matching phase distributions are shown in the bottom panels (\textbf{b}). With trapping frequencies \( \omega_x = 10.0 \) and \( \omega_y = 10.0 \), the system is contained within a isotropic harmonic trap. The following interaction parameters are used: interspecies interaction \( g_{12} = 10.0 \) and intraspecies interaction \( g_1 = g_2 = 100.0 \).}
    \label{fig:12}
\end{figure}
As observed in Fig.12, we have considered harmonic potential with trap frequencies along both \(x\) and \(y\) set as \(\omega_x = \omega_y = 10.0\). This ensures an isotropic trap. The helicoidal spin-orbit coupling strength has been set at \(\alpha = 1\). The intra-species interaction parameters have been fixed at \(g_1 = g_2 = 100.0\), with the inter-component term as \(g_{12} = 10.0\). This ensures that the system is in a regime conducive to inter-component binding. The helicoidal spin-orbit coupling has equal contributions from the Rabi-coupling term R as well. The condensate is initialized with a symmetric Gaussian profile, with both the components \(\psi_1\) and \(\psi_2\) allowed to evolve under the Gross-Pitaevskii framework. As shown in Fig.12, the spatial density profiles \(|\psi_1|^2\) and \(|\psi_2|^2\) of the condensate evolve into ring-like structures, exhibiting radial modulation with a well-pronounced central density dip which are surrounded by lobed structures. These features emerge due to the interplay between the repulsive interactions and the helicoidal spin-orbit coupling, redistributing density non-uniformly across space. The total density \(|\psi_1|^2 + |\psi_2|^2\) shows a cylindrical symmetry that remains consistent with the isotropic trap. This confirms the spatial coherence maintained by both components which share a similar  morphology. The phase profiles reveal multiple phase windings and vortex-like defects symmetrically distributed around the center. Notably, the phase of the combined field \((|\psi_1|^{2} + |\psi_2|^{2})\) displays structured interference patterns and is quite indicative of coherent coupling. This configuration serves as a reference case for understanding the symmetric dynamical behavior of the condensate under uniform harmonic confinement, helicoidal spin–orbit coupling, and Rabi coupling.

\subsection*{Case 2: Anisotropic Trap: \(\omega_x > \omega_y\)}
\begin{figure}[h]
    \centering
    \xincludegraphics[width=1\linewidth,label=\textbf{a.}]{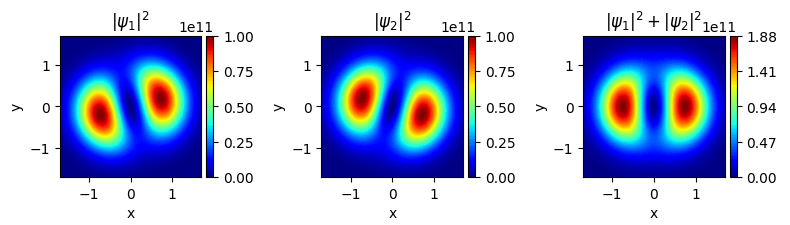}
    \xincludegraphics[width=1\linewidth,label=\textbf{b.}]{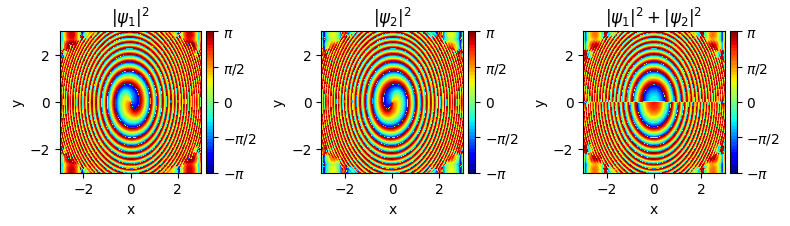}
    \caption{The spatial density distributions of the two-component wavefunctions \( \psi_1(x,y) \), \( \psi_2(x,y) \), and their total density \( |\psi_1|^2 + |\psi_2|^2 \) are displayed in the top panels (\textbf{a}). The matching phase distributions are shown in the bottom panels (\textbf{b}). With trapping frequencies \( \omega_x = 48.0 \) and \( \omega_y = 33.0 \), the system is contained within an anisotropic harmonic trap. The following interaction parameters are used: interspecies interaction \( g_{12} = 10.0 \) and intraspecies interaction \( g_1 = g_2 = 100.0 \).}
    \label{fig:1}
\end{figure}
In this case, as observed in Fig. 13, we have introduced an anisotropic trapping potential with the interaction parameters kept at the same values as in the previous case. The trapping frequencies have been changed to \(\omega_x = 48.0\) and \(\omega_y = 33.0\) for this case. This tightens the confinement along the x-direction. The spatial density profiles in Fig. 13 show an elongation of both components relatively more along the x-axis, the axis with stronger confinement. Compared to the isotropic case, the lobes of \(|\psi_1|^2\) and \(|\psi_2|^2\) are now spatially extended and shifted along the direction of stronger trapping frequency, while the total density remains localized, completely along x-axis. The condensate’s elongation and deformation are direct consequences of the anisotropic confinement combined with the interplay of helicoidal spin–orbit coupling and Rabi coupling. The phase plots reflect some asymmetry as well, revealing slight vortex displacement and asymmetrically distributed phase windings. The combined phase structure \((|\psi_1|^{2} + |\psi_2|^{2})\) shows non-uniform gradients, suggesting altered interference behavior due to spatial anisotropy. Thus, this simulation underscores the role of trap geometry in shaping the condensate structure and offers insight into the control of condensate configurations via external potentials.

\subsection*{Case 3: Anisotropic Trap: \(\omega_x < \omega_y\)}
\begin{figure}[h]
    \centering
    \xincludegraphics[width=1\linewidth,label=\textbf{a.}]{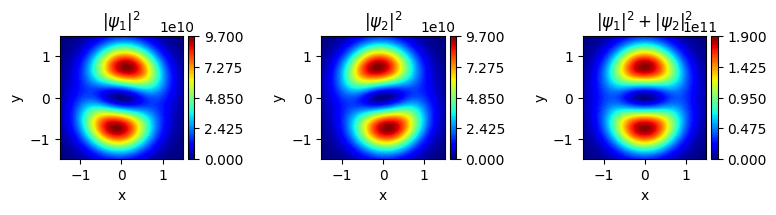}
    \xincludegraphics[width=1\linewidth,label=\textbf{b.}]{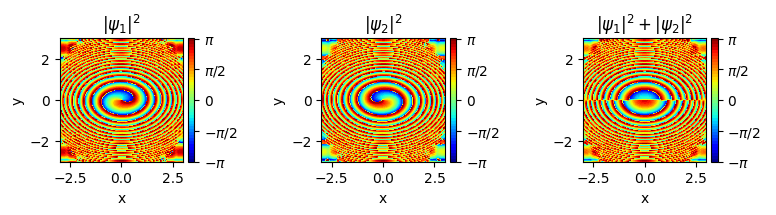}
    \caption{The spatial density distributions of the two-component wavefunctions \( \psi_1(x,y) \), \( \psi_2(x,y) \), and their total density \( |\psi_1|^2 + |\psi_2|^2 \) are displayed in the top panels (\textbf{a}). The matching phase distributions are shown in the bottom panels (\textbf{b}). With trapping frequencies \( \omega_x = 33.0 \) and \( \omega_y = 48.0 \), the system is contained within an anisotropic harmonic trap. The following interaction parameters are used: interspecies interaction \( g_{12} = 10.0 \) and intraspecies interaction \( g_1 = g_2 = 100.0 \).}
    \end{figure}
In contrast to the previous configuration, we have now considered an inverse anisotropy in the trapping potential by setting \(\omega_x = 33.0\) and \(\omega_y = 48.0\). All the other interaction parameters remain unchanged in this case as well. This flip in confinement symmetry rotates the direction of tighter spatial confinement from the x-axis to the y-axis. The density plots (see Fig. 14) reveal an apparent elongation of both components along the y-direction, orthogonal to the weaker x-axis confinement. In contrast to the previously discussed case, the spatial modulation is now predominantly distributed along the y-axis, with wave-fronts forming ripple-like textures due to the interference effects induced by the coupling terms. The individual densities \(|\psi_1|^2\) and \(|\psi_2|^2\) remain overlapping but continue to exhibit asymmetries in their inner core structures. This confirms non-linear self-organization taking place within the anisotropic trap. The total density again shows strong localization with a more prominent lobed profile along the axis with a higher value of trapping frequency. The phase plot of \(\psi_1 + \psi_2\) exhibits flattened contours. This is clearly indicative at constructive interference patterns emerging from the directional asymmetry. This simulation clearly underlines that not only the strength of the trap but also its directionality plays a key role in shaping the condensate dynamics. As seen in this case, flipped anisotropy produces distinctly rotated yet stable condensate configurations. Thus, reaffirming the controllability of such systems.

\section{Conclusion}

To summarise, a comprehensive modulation instability (MI) gain analysis was performed for a two-component Bose–Einstein condensate (BEC) with helicoidal SOC and Rabi-coupling under a two-dimensional harmonic potential. Our main aim was to determine how the in-terplay between nonlinear atomic interactions and spin
orbit-induced coherence affects the stability of the condensate. Through the exploration of the MI gain for a range of different parameters, we have outlined the specific roles played by inter- and intra-species interactions, HSOC strength, and Rabi coupling in controlling the onset and development of the instability. The analysis reveals that increasing both the HSOC strength ($\alpha$) and Rabi coupling ($R$) leads to higher MI gain accompanied with an expansion of instability regions in the momentum space, especially in the high-momentum ($|k_x|$, $|k_y|$) domain. This brings forth the system’s enhanced susceptibility to density perturbations, a critical factor in the emergence of localized condensate structures and dynamical instabilities. However, strong intra-species repulsion ($g_1$, $g_2$) tends to suppress MI and narrows the instability bandwidth. The inter-species interaction ($g_{12}$) is found to be pivotal, particularly in the attractive regime as it not only amplifies MI gain but also broadens the momentum-space regions over which the instability sustains itself. Furthermore, a joint analysis by varying both $g_{12}$ and $R$ together revealed a mutual enhancement of MI gain when both parameters are large. This highlights a crucial interplay between atomic interaction strength and coherent spin
mixing by suggesting that the simultaneous tuning of these parameters creates optimal conditions for stable condensate. Conversely, the MI is observed to be suppressed significantly where only $R$ is highly negative. Overall, our findings provide a clear blueprint for multiple combinations of parameter spaces that promote or hinder modulation instability. By means of numerical simulations on the coupled Gross–Pitaevskii equations with both helicoidal spin–orbit and Rabi coupling, we have demonstrated the formation and development of stable condensate structures in two-dimensional harmonic traps. The condensate geometry is largely controlled by the anisotropy in confinement, whereas phase structures show interference and coherence caused by spin-orbit effects. Thus, this work represents a major step toward controlled pattern formation in spin–orbit–coupled BECs by tying the analysis of modulation instability gain to a more comprehensive understanding of condensate dynamics.

\section{Acknowledgment}
S. Saravana Veni acknowledges
Amrita Vishwa Vidyapeetham, Coimbatore where this work was supported under Amrita Seed Grant (File Number: ASG2022141).

\section{Appendix}
With reference to the matrix in equation~8 of this paper, please find below the fully expanded version with all the matrix elements. The term \(\lambda = \frac{1}{2}( -k_x^2 - k_y^2 ) - \frac{1}{2}(x^2 + y^2)\omega^2 \) :

\begin{widetext}
\begin{equation}
M = 
\begin{pmatrix}
\beta(k_x + k_y) + \Omega & -\alpha(k_x + k_y) & \lambda & -\frac{R}{2} \\
-\alpha(k_x + k_y) & \beta(k_x + k_y) + \Omega & -\frac{R}{2} & \lambda \\
-2g_1n_{10} + \lambda & -2g_{12}\sqrt{n_{10}n_{20}} - R & \beta(k_x + k_y) + \Omega & -\alpha(k_x + k_y) \\
-2g_{12}\sqrt{n_{10}n_{20}} - R & -2g_2n_{20} + \lambda & -\alpha(k_x + k_y) & \beta(k_x + k_y) + \Omega
\end{pmatrix}
\end{equation}

The coefficient terms with respect to equation~9, \(P_j\) (j =1, 2, 3) have been mentioned below:
\footnotesize
\begin{align}
P_1 &= - (k_x + k_y) \bigg( 
4 \alpha^2 \beta (k_x + k_y)^2 - 4 \beta^3 (k_x + k_y)^2 
+ 4 \alpha g_{12} (k_x^2 + k_y^2) \sqrt{n_{10} n_{20}} \notag \\
&\quad + \beta (k_x^2 + k_y^2) \left(k_x^2 + k_y^2 + 2 g_1 n_{10} + 2 g_2 n_{20} \right) 
+ \alpha (3 k_x^2 + 3 k_y^2 + 2 g_1 n_{10} + 2 g_2 n_{20}) R \notag \\
&\quad + 2 \beta R (2 g_{12} \sqrt{n_{10} n_{20}} + R) 
\bigg), \\
P_2 &= -2 \alpha^2 (k_x + k_y)^2 + 6 \beta^2 (k_x + k_y)^2 
- \frac{1}{2} (k_x^2 + k_y^2) \left(k_x^2 + k_y^2 + 2 g_1 n_{10} + 2 g_2 n_{20} \right) 
- R \left(2 g_{12} \sqrt{n_{10} n_{20}} + R \right), \\
P_3 &= 4 \beta (k_x + k_y)
\end{align}
\end{widetext}

\section{References}

\bibliographystyle{unsrt}

\bibliography{references}

@article{ref1,
  author    = {V.~E. Zakharov and L.~A. Ostrovsky},
  title     = {Modulation instability: The beginning},
  journal   = {Physica D},
  volume    = {238},
  pages     = {540--548},
  year      = {2009}
}

@article{ref2,
  author    = {Y.~V. Kartashov and V.~V. Konotop and M. Modugno and E.~Y. Sherman},
  title     = {Solitons in Inhomogeneous Gauge Potentials: Integrable and Nonintegrable Dynamics},
  journal   = {Phys. Rev. Lett.},
  volume    = {122},
  pages     = {064101},
  year      = {2019}
}

@article{ref3,
  author    = {{Y.~V. Kartashov and V.~V. Konotop}},
  title     = {{Solitons in Bose-Einstein Condensates with Helicoidal Spin-Orbit Coupling}},
  journal   = {{Phys.\ Rev.\ Lett.}},
  volume    = {{118}},
  pages     = {{190401}},
  year      = {{2017}}
}

@article{ref4,
  author    = {E.~V. Goldstein and P. Meystre},
  title     = {Dipole-dipole interaction in optical cavities},
  journal   = {Phys. Rev. A},
  volume    = {56},
  pages     = {5135},
  year      = {1997}
}

@article{ref5,
  author    = {E.~V. Goldstein and P. Meystre},
  title     = {Quasiparticle instabilities in multicomponent atomic condensates},
  journal   = {Phys. Rev. A},
  volume    = {55},
  pages     = {2935},
  year      = {1997}
}

@article{ref6,
  author    = {Wen-Rong Sun and Jin-Hua Li and Lei Liu and P.G. Kevrekidis},
  title     = {The instabilities beyond modulational type in a repulsive Bose–Einstein condensate with a periodic potential},
  journal   = {Physica D: Nonlinear Phenomena},
  volume    = {458},
  pages     = {134009},
  year      = {2024}
}

@article{ref7,
  author    = {Theocharis, G. and Schmelcher, P. and Kevrekidis, P. G. and Frantzeskakis, D. J.},
  title     = {Matter-wave solitons of collisionally inhomogeneous condensates},
  journal   = {Phys. Rev. A},
  volume    = {72},
  pages     = {033614},
  year      = {2005}
}

@article{ref8,
  author    = {Salasnich, L. and Parola, A. and Reatto, L.},
  title     = {Modulational Instability and Complex Dynamics of Confined Matter-Wave Solitons},
  journal   = {Phys. Rev. Lett.},
  volume    = {91},
  pages     = {080405},
  year      = {2003}
}

@article{ref9,
  author    = {Maura Brunetti and Jérôme Kasparian},
  title     = {Modulational instability in wind-forced waves},
  journal   = {Phys. Lett. A},
  volume    = {378},
  pages     = {3626-3630},
  year      = {2014}
}

@article{ref10,
  author    = {Gu, Qiang and Qiu, Haibo},
  title     = {Coherent Dynamics of Domain Formation in the Bose Ferromagnet},
  journal   = {Phys. Rev. Lett.},
  volume    = {98},
  pages     = {200401},
  year      = {2007}
}

@article{ref11,
  author    = {Goldman, Martin V.},
  title     = {Strong turbulence of plasma waves},
  journal   = {Rev. Mod. Phys.},
  volume    = {56},
  pages     = {709--735},
  year      = {1984}
}

@article{ref12,
  author    = {Wen, Wen and Huang, Guoxiang},
  title     = {Dynamics of dark solitons in superfluid Fermi gases in the BCS-BEC crossover},
  journal   = {Phys. Rev. A},
  volume    = {79},
  pages     = {023605},
  year      = {2009}
}

@article{ref13,
  author    = {Yu, Xiaoquan and Blakie, P. B.},
  title     = {Propagating Ferrodark Solitons in a Superfluid: Exact Solutions and Anomalous Dynamics},
  journal   = {Phys. Rev. Lett.},
  volume    = {128},
  pages     = {125301},
  year      = {2022}
}

@article{ref14,
  author    = {Flores-Calderón, R. and Fujioka, J. and Espinosa-Cerón, A.},
  title     = {Soliton dynamics of a high-density Bose-Einstein condensate subject to a time varying anharmonic trap},
  journal   = {Chaos, Solitons \& Fractals},
  volume    = {143},
  pages     = {110580},
  year      = {2021}
}

@article{ref15,
  author    = {Stanescu, Tudor D. and Anderson, Brandon and Galitski, Victor},
  title     = {Spin-orbit coupled Bose-Einstein condensates},
  journal   = {Phys. Rev. A},
  volume    = {78},
  pages     = {023616},
  year      = {2008}
}

@article{ref16,
  author    = {{Kato YK, Myers RC, Gossard AC, Awschalom DD.}},
  title     = {Observation of the spin Hall effect in semiconductors},
  journal   = {Science},
  volume    = {306},
  pages     = {1910-1913},
  year      = {2004}
}

@article{ref17,
  author    = {Hasan, M. Z. and Kane, C. L.},
  title     = {Colloquium: Topological insulators},
  journal   = {Rev. Mod. Phys.},
  volume    = {82},
  pages     = {3045--3067},
  year      = {2010}
}

@article{ref18,
  author    = {Schneider, M. et al.},
  title     = {Control of the Bose-Einstein Condensation of Magnons by the Spin Hall Effect},
  journal   = {Phys. Rev. Lett.},
  volume    = {127},
  pages     = {237203},
  year      = {2021}
}

@article{ref19,
  author    = {Divinskiy, B. et al.},
  title     = {Evidence for spin current driven Bose--Einstein condensation of magnons},
  journal   = {Nat. Commun.},
  volume    = {12},
  pages     = {6541},
  year      = {2021}
}

@article{ref20,
  author    = {Otajonov, Sherzod R. and Tsoy, Eduard N. and Abdullaev, Fatkhulla Kh.},
  title     = {Modulational instability and quantum droplets in a two-dimensional Bose-Einstein condensate},
  journal   = {Phys. Rev. A},
  volume    = {106},
  pages     = {033309},
  year      = {2022}
}

@article{ref21,
  author    = {Otajonov, Sherzod R. and Umarov, Bakhram A. and Abdullaev, Fatkhulla Kh.},
  title     = {Modulational instability and discrete quantum droplets in a deep quasi-one-dimensional optical lattice},
  journal   = {Phys. Rev. E},
  volume    = {111},
  pages     = {054206},
  year      = {2025}
}

@article{ref22,
  title={Anderson localization of a non-interacting Bose--Einstein condensate},
  author={Roati, Giacomo and D’Errico, Chiara and Fallani, Leonardo and Fattori, Marco and Fort, Chiara and Zaccanti, Matteo and Modugno, Giovanni and Modugno, Michele and Inguscio, Massimo},
  journal={Nature},
  volume={453},
  number={7197},
  pages={895--898},
  year={2008},
  publisher={Nature Publishing Group UK London}
}

@article{ref23,
  author    = {Saito, Hiroki and Ueda, Masahito},
  title     = {Mean-field analysis of collapsing and exploding Bose-Einstein condensates},
  journal   = {Phys. Rev. A},
  volume    = {65},
  pages     = {033624},
  year      = {2002}
}

@article{ref24,
  author    = {Kevrekidis, P. G. and Frantzeskakis, D. J. and Carretero-Gonz{\'a}lez, R.},
  title     = {Basic Mean-Field Theory for Bose-Einstein Condensates},
  publisher   = {Springer Berlin Heidelberg},
  volume    = {45},
  pages     = {3--21},
  year      = {2008}
}

@article{ref25,
  author    = {Zhang, Yongping and Mao, Li and Zhang, Chuanwei},
  title     = {Mean-Field Dynamics of Spin-Orbit Coupled Bose-Einstein Condensates},
  journal   = {Phys. Rev. Lett.},
  volume = {108},
  pages = {035302},
  year      = {2012}
}

@article{ref26,
  author    = {W Ketterle},
  title     = {Bose–Einstein condensation in dilute atomic gases: atomic physics meets condensed matter physics},
  journal   = {Physica B: Condensed Matter},
  volume    = {280},
  pages     = {11--19},
  year      = {2000}
}

@article{ref27,
  author    = {{Anglin J., Ketterle W.}},
  title     = {Bose–Einstein condensation of atomic gases},
  journal   = {Nature},
  volume    = {416},
  pages     = {211-218},
  year      = {2002}
}

@article{ref28,
  author    = {Li, Xiao-Xun and Cheng, Rui-Jin and Zhang, Ai-Xia and Xue, Ju-Kui},
  title     = {Modulational instability of Bose-Einstein condensates with helicoidal spin-orbit coupling},
  journal   = {Phys. Rev. E},
  volume    = {100},
  pages     = {032220},
  year      = {2019}
}

@article{ref29,
  author    = {Kong, C. and Yin, B. and Wu, J. and Huang, J. et al.},
  title     = {Stability control in a helicoidal spin–orbit-coupled open Bose–Bose mixture},
  journal   = {Open Physics},
  volume    = {21},
  pages     = {0263},
  year      = {2022}
}

@article{ref30,
  author    = {Sobkowiak, S. and Semkat, D. and Stolz, H. and Koch, Th. and Fehske, H.},
  title     = {Interacting multicomponent exciton gases in a potential trap: Phase separation and Bose-Einstein condensation},
  journal   = {Phys. Rev. B},
  volume    = {82},
  pages     = {064505},
  year      = {2010}
}

@article{ref31,
  author    = {Gaunt, Alexander L. and Schmidutz, Tobias F. and Gotlibovych, Igor and Smith, Robert P. and Hadzibabic, Zoran},
  title     = {Bose-Einstein Condensation of Atoms in a Uniform Potential},
  journal   = {Phys. Rev. Lett.},
  volume    = {110},
  pages     = {200406},
  year      = {2013}
}

@article{ref32,
  author    = {Bloch, J. and Carusotto, I. et al.},
  title     = {Non-equilibrium Bose--Einstein condensation in photonic systems},
  journal   = {Nat. Rev. Phys.},
  volume    = {4},
  pages     = {470--488},
  year      = {2022}
}

@article{ref33,
  author    = {H. Huang and H. Wang and C. S. Lim and K.-C. Wong},
  title     = {Anisotropic dipolar vortex quantum droplets in an annular potential},
  journal   = {Chaos, Solitons \& Fractals},
  volume    = {190},
  pages     = {115762},
  year      = {2025},
  doi       = {10.1016/j.chaos.2024.115762}
}

@article{ref34,
  author    = {M. Chen and B. Liao and H. Wang and H. Huang et al. },
  title     = {Composite solitons in spatially confined spin-orbit-coupled BEC with dipole-dipole repulsion},
  journal   = {Results Phys.},
  volume    = {18},
  pages     = {103304},
  year      = {2020},
  doi       = {10.1016/j.rinp.2020.103304}
}

@article{ref35,
  author    = {D.~L. Campbell and G. Juzeli{\=u}nas and I.~B. Spielman},
  title     = {Realistic Rashba and Dresselhaus spin--orbit coupling for neutral atoms},
  journal   = {Phys. Rev. A},
  volume    = {84},
  pages     = {025602},
  year      = {2011}
}

@article{ref36,
  author    = {B.~M. Anderson and G. Juzeli{\=u}nas and N.~R. Cooper},
  title     = {Synthetic 3D spin--orbit coupling},
  journal   = {Phys. Rev. Lett.},
  volume    = {108},
  pages     = {235301},
  year      = {2012}
}

@article{ref37,
  author    = {V. Galitski and I.~B. Spielman},
  title     = {Spin--orbit coupling in quantum gases},
  journal   = {Nature},
  volume    = {494},
  pages     = {49--54},
  year      = {2013}
}

@article{ref38,
  author    = {J.-R. Li and J. Lee and W. Huang and S. Burchesky and B. Shteynas and F. {\c{C}}. Top and A.~O. Jamison and W. Ketterle},
  title     = {A stripe phase with supersolid properties in spin--orbit-coupled Bose--Einstein condensates},
  journal   = {Nature},
  volume    = {543},
  pages     = {91--94},
  year      = {2017}
}

@article{ref39,
  author    = {Neslihan Uzar, Sedat Ballikaya},
  title     = {Investigation of classical and fractional Bose–Einstein condensation for harmonic potential},
  journal   = {Physica A: Statistical Mechanics and its Applications},
  volume    = {392},
  pages     = {1733-1741},
  year      = {2013}
}

@article{ref40,
  author    = {Berry, N. Hawk and Kutz, J. Nathan},
  title     = {Dynamics of Bose-Einstein condensates under the influence of periodic and harmonic potentials},
  journal   = {Phys. Rev. E},
  volume    = {75},
  pages     = {036214},
  year      = {2007}
}

@article{ref41,
  author    = {Zhang, Xiao-Fei and Hu, Xing-Hua and Liu, Xun-Xu and Liu, W. M.},
  title     = {Vector solitons in two-component Bose-Einstein condensates with tunable interactions and harmonic potential},
  journal   = {Phys. Rev. A},
  volume    = {79},
  pages     = {033630},
  year      = {2009}
}

@article{ref42,
  author    = {Berman, Oleg L. and Lozovik, Yurii E. and Snoke, David W.},
  title     = {Theory of Bose-Einstein condensation and superfluidity of two-dimensional polaritons in an in-plane harmonic potential},
  journal   = {Phys. Rev. B},
  volume    = {77},
  pages     = {155317},
  year      = {2008}
}

@article{ref43,
  title={The two-dimensional Bose--Einstein condensate},
  author={Fern{\'a}ndez, Juan Pablo and Mullin, William J},
  journal={J. Low Temp. Phys.},
  volume={128},
  number={5},
  pages={233--249},
  year={2002},
  publisher={Springer}
}

@article{ref44,
  author    = {S Bhuvaneswari et al},
  title     = {Modulation instability in quasi-two-dimensional spin–orbit coupled Bose–Einstein condensates},
  journal   = {J. Phys. B},
  volume    = {49},
  pages     = {245301},
  year      = {2016}
}

@article{ref45,
  author    = {Conrad Bertrand Tabi and Phelo Otladisa and Timoléon Crépin Kofané},
  title     = {Modulation instability of two-dimensional Bose--Einstein condensates with helicoidal and a mixture of Rashba--Dresselhaus spin--orbit couplings},
  journal   = {Phys. Lett. A},
  volume    = {449},
  pages     = {128334},
  year      = {2022}
}

@article{ref46,
title = {Vortex droplets and lattice patterns in two-dimensional traps: A photonic spin–orbit-coupling perspective},
journal = {Chaos, Solitons \& Fractals},
volume = {197},
pages = {116441},
year = {2025},
author = {S. Sanjay and S. Saravana Veni and Boris A. Malomed},
}

@article{ref47,
title = {Synergistic effects of spin-orbit coupling and intercomponent interactions in two-component (2+1)D photonic fields},
journal = {Chaos, Solitons \& Fractals},
volume = {199},
pages = {116806},
year = {2025},
author = {Suri Deekshita and S. Sanjay and S. Saravana Veni and Conrad B. Tabi and Timoléon C. Kofané},
}

@article{rechtsman2013photonic,
  title={{Photonic Floquet topological insulators}},
  author={{Rechtsman, Mikael C and Zeuner, Julia M and Plotnik, Yonatan and Lumer, Yaakov and Podolsky, Daniel and Dreisow, Felix and Nolte, Stefan and Segev, Mordechai and Szameit, Alexander}},
  journal={{Nature}},
  volume={{496}},
  number={{7444}},
  pages={{196--200}},
  year={{2013}},
  publisher={Nature Publishing Group UK London}
}

@article{lin2011spin,
  title={{Spin--orbit-coupled Bose--Einstein condensates}},
  author={{Y.-J. Lin and K. Jim{\'e}nez-Garc{\'\i}a and I. B. Spielman}},
  journal={{Nature}},
  volume={{471}},
  number={{7336}},
  pages={{83--86}},
  year={{2011}},
  publisher={Nature Publishing Group UK London}
}

@article{otlaadisa2021modulation,
  title={{Modulation instability in helicoidal spin-orbit coupled open Bose-Einstein condensates}},
  author={{P. Otlaadisa and C. B. Tabi and T. C. Kofan{\'e}}},
  journal={{Phys.\ Rev.\ E}},
  volume={{103}},
  number={{5}},
  pages={{052206}},
  year={{2021}},
  publisher={APS}
}

\end{document}